\documentclass[twocolumn,trackchanges]{aastex701} 
\usepackage{amsmath}

\begin{document}

\title{A Catalog of 971 FR-I Radio Galaxies from the FIRST Survey via Hybrid Deep Learning and Ridgeline Flux Density Distribution Analysis}

\author[orcid=0000-0002-3426-3269,sname='Lao']{Baoqiang Lao}
\affiliation{Guangxi Key Laboratory of Brain-inspired Computing and Intelligent Chips, School of Electronic and Information Engineering, GuangXi Normal University, Guilin 541004, People's Republic of China}
\affiliation{Key Laboratory of Nonlinear Circuits and Optical Communications (GuangXi Normal University), Education Department of Guangxi Zhuang Autonomous Region, Guilin 541004, People's Republic of China}
\email[show]{lbq@gxnu.edu.cn}  

\author[orcid=0000-0002-4439-5580,sname='Yang']{Xiaolong Yang}
\affiliation{Shanghai Astronomical Observatory, Chinese Academy of Sciences, Shanghai 200030, People's Republic of China}
\email[show]{yangxl@shao.ac.cn}

\author[sname='Zhao']{Wenjun Xiao}
\affiliation{School of Physics and Electronic Science, Guizhou Normal University, Guiyang 550001, People's Republic of China}
\email{xiaowenjun@gznu.edu.cn}

\author[orcid=0000-0001-9251-9456,gname=Kumar,sname='Sasmal']{Tapan K. Sasmal} 
\affiliation{National Astronomical Observatories, Chinese Academy of Sciences, Beijing 100101, People's Republic of China}
\email{tapankumar@bao.ac.cn}

\author[sname='Liu']{Yanli Zou}
\affiliation{Guangxi Key Laboratory of Brain-inspired Computing and Intelligent Chips, School of Electronic and Information Engineering, GuangXi Normal University, Guilin 541004, People's Republic of China}
\affiliation{Key Laboratory of Nonlinear Circuits and Optical Communications (GuangXi Normal University), Education Department of Guangxi Zhuang Autonomous Region, Guilin 541004, People's Republic of China}
\email{eeyzou@gxnu.edu.cn}

\author[sname='Liu']{Didi Liu}
\affiliation{Guangxi Key Laboratory of Brain-inspired Computing and Intelligent Chips, School of Electronic and Information Engineering, GuangXi Normal University, Guilin 541004, People's Republic of China}
\affiliation{Key Laboratory of Nonlinear Circuits and Optical Communications (GuangXi Normal University), Education Department of Guangxi Zhuang Autonomous Region, Guilin 541004, People's Republic of China}
\email{ldd866@gxnu.edu.cn}

\author[sname='Liao']{Zhixian Liao}
\affiliation{Guangxi Key Laboratory of Brain-inspired Computing and Intelligent Chips, School of Electronic and Information Engineering, GuangXi Normal University, Guilin 541004, People's Republic of China}
\affiliation{Key Laboratory of Nonlinear Circuits and Optical Communications (GuangXi Normal University), Education Department of Guangxi Zhuang Autonomous Region, Guilin 541004, People's Republic of China}
\email{zhixianliao@gxnu.edu.cn}

\author[sname='Lu']{Ye Lu}
\affiliation{Guangxi Key Laboratory of Brain-inspired Computing and Intelligent Chips, School of Electronic and Information Engineering, GuangXi Normal University, Guilin 541004, People's Republic of China}
\affiliation{Key Laboratory of Nonlinear Circuits and Optical Communications (GuangXi Normal University), Education Department of Guangxi Zhuang Autonomous Region, Guilin 541004, People's Republic of China}
\email{luye@gxnu.edu.cn}

\author[orcid=0000-0002-1243-0476, sname='Zhao']{Rushuang Zhao}
\affiliation{School of Physics and Electronic Science, Guizhou Normal University, Guiyang 550001, People's Republic of China}
\email{201907007@gznu.edu.cn}



\begin{abstract}

We present a catalog of 971 FR-I radio galaxies (FR-Is) identified from the Very Large Array Faint Images of the Radio Sky at Twenty-Centimeters (FIRST) survey. The identifications were made using a hybrid method that combines deep learning with ridgeline flux density distribution analysis. Among these sources, 845 are new discoveries. The catalog comprises sources characterized by edge-darkened double jets, an absence of significant bent morphology, and angular sizes ranging from 23 to 159 arcseconds. Optical and/or infrared counterparts have been identified for 813 FR-Is. Among these, the host galaxies are predominantly (88.1\%) red galaxies, with the remainder (11.9\%) being blue galaxies; notably, most blue galaxies exhibit high radio power. 
The FR-I sample spans a radio power range of $1.20 \times 10^{21} \leq P_{\rm 1400} \leq 3.55 \times 10^{27} \, {\rm W\,Hz}^{-1}$ at 1400 MHz and reaches redshifts up to $z = 2.307$. The host galaxies have $r$-band absolute magnitudes in the range $-24 \lesssim M_r \lesssim -20$ mag. For the 512 FR-Is with estimates, the black hole masses fall within $10^7 \lesssim M_{\rm BH} \lesssim 7.94 \times 10^9 \, M_{\odot}$. Based on optical emission-line ratios and mid-infrared colors, spectroscopic classification shows that 571 hosts are low-excitation radio galaxies (LERGs) and 59 are high-excitation radio galaxies (HERGs). 


\end{abstract}




\section{Introduction} \label{Sec:Intro}
Radio galaxies represent prominent active galactic nuclei (AGN) phenomena powered by supermassive black holes (SMBHs). The dynamical characteristics of AGN jets and their interactions with the intergalactic medium result in a continuous distribution of radio morphologies. In terms of spatial scales, they range from parsec (pc)-scale compact sources (CS) and core-jet (CJ) structures to megaparsec (Mpc)-scale extended double lobes. Morphologically, they encompass the classical Fanaroff-Riley (FR) type I (FR-I) and type II (FR-II) classifications \citep{1974MNRAS.167P..31F}, along with dynamically disturbed structures such as bent-tail (BT) sources \citep[e.g.,][]{2025ApJS..276...46L} and peculiar X/S/Z-shaped morphologies \citep[e.g.,][]{2022MNRAS.512.4308B,2025MNRAS.536.2025M}. The diversity of radio morphologies directly reflects the physical processes and evolutionary stages of jet-medium interactions. This research holds critical significance for addressing major astronomical questions, including galaxy formation and evolutionary history \citep[e.g.,][]{2025A&A...695A.178V}, SMBH-host galaxy co-evolution \citep[e.g.,][]{2023NatAs...7.1376Z}, and AGN feedback mechanisms \citep[e.g.,][]{2023NatAs...7.1275A}. Furthermore, it offers a unique probe of extreme astrophysical phenomena, including ultra-large-scale ($\sim$7 Mpc) jets \citep{2024Natur.633..537O} and odd radio circles (ORCs) \citep{2024Natur.625..459C}.

Among them, FR-I radio galaxies (FR-Is), characterized by their edge-darkened radio morphology and trans-relativistic jets \citep{1974MNRAS.167P..31F}, serve as critical laboratories for probing jet-interstellar medium (jet-ISM) interactions \citep[e.g.,][]{2015A&A...574A..89S} and the energetics of low-power AGN \citep{2009ApJ...696.1103C}. Despite their prevalence, systematic studies of FR-Is remain challenging, hindered by contamination from unresolved sources, projection effects, and the scarcity of uniformly selected samples \citep[e.g.,][]{2017AA...598A..49C}.

Beyond morphological classification, radio galaxies can be spectroscopically divided into high-excitation radio galaxies (HERGs) and low-excitation radio galaxies (LERGs) based on emission-line ratios (e.g., [OIII]$\lambda$5007/${\rm H}_{\alpha}$) according to the excitation scheme of \citet{1994ASPC...54..201L}. Observations consistently show that FR-Is are predominantly LERGs \citep{2009ApJ...699L..43S,2012MNRAS.421.1569B}, characterized by weak or absent high-ionization lines, which indicates radiatively inefficient accretion from hot gas reservoir \citep{2016MNRAS.460....2P}. In contrast, FR-IIs are typically HERGs associated with efficient accretion of cold gas accretion. The FR-I/FR-II dichotomy thus reflects distinct accretion regimes: FR-Is typically exhibit jet-dominated morphology powered by a sub-Eddington advection-dominated accretion flow, whereas FR-IIs sustain radiatively efficient disks through near-Eddington accretion \citep{2014ARA&A..52..589H}. 



The advent of large-area radio surveys, such as the Very Large Array (VLA) Sky Survey \citep[VLASS;][]{2020PASP..132c5001L} and the VLA Faint Images of the Radio Sky at Twenty-Centimeters \citep[FIRST;][]{1995ApJ...450..559B,1997ApJ...475..479W} survey, has revolutionized statistical analyses of radio galaxies. However, morphological classification still relies heavily on time-consuming and subjective visual inspection \citep{2025MNRAS.536.3488W}. While recent deep learning approaches to automate this task have shown great promise \citep[e.g.,][]{2021SciBu..66.2145L,2021MNRAS.501.4579B,2023A&C....4400728L,2024PASA...41....1G}, residual false positives necessitate hybrid methods that combine artificial intelligence with human vetting \citep{2025ApJS..276...46L}.

In this study, we present a catalog of 971 FR-Is identified through a two-stage pipeline: (1) a deep learning-based detector applied to FIRST images, followed by (2) the ridgeline flux density distribution-base method to remove misclassifications. Our sample, an order of magnitude larger than those in previous works \citep[e.g.,][]{2017AA...598A..49C}, enables a statistically robust exploration of the FR-I properties. We further analyze their multi-wavelength characteristics, linking radio morphology to optical host properties. This paper is structured as follows: Section~\ref{Sec:sample} describes the data and methodology; Section~\ref{Sec:RE} presents the FR-I catalog and discusses the results; Section~\ref{Sec:Conc} provides the conclusions.
Throughout this paper, we adopt a flat $\Lambda$CDM cosmological model with $H_0 = 70 \,{\rm km}\,{\rm s}^{-1}\,{\rm Mpc}^{-1}$ and $\Omega_m = 0.3$. 

\section{Sample selection and characterization}\label{Sec:sample}
\subsection{Data source: VLA FIRST survey}
The FIRST survey was conducted using the VLA in its B-configuration, observing at 1400~MHz (L-band) with a bandwidth of 3 MHz per channel. The survey covered 10,575~deg$^2$ in both the North and South Galactic Cap regions, achieving a typical root-mean-square (RMS) noise of 0.15 mJy beam$^{-1}$ and positional accuracy better than 1 arcsec for sources above 3 mJy. The angular resolution reached 5.4 arcsec full width at half-maximum (FWHM) with circular restoration beams in the Northern sky, whereas the Southern sky exhibited slightly coarser resolution with elliptical beams, measuring 6.4$\times$5.4 arcsec FWHM, and in some cases up to 6.8$\times$5.4 arcsec FWHM. Our analysis utilizes the 17 December 2014 data release \citep{2015ApJ...801...26H}, which provides a catalog of 946,432 components with peak flux densities $\geq$1~mJy and co-added Stokes I maps with a 1.8 arcsec pixel scale.

\subsection{Automated morphological classification and ridgeline flux density distribution-based identification method}
The selection began with an automated classification via the deep learning-based Radio Galaxy Classification with Mask Transfiner \citep[RGCMT;][]{2023A&C....4400728L} which identified and categorized FIRST radio sources into five morphological types: CS, FR-I, FR-II, BT, and CJ. The RGCMT model generated a predicted mask, bounding box, and confidence score for each source. The centroid position of each source, calculated from its predicted mask, will be used in post-processing. Each FR-I's predicted mask was generated by applying a 3$\times$local-RMS threshold to the source's bounding box in the FIRST image. As reported in our previous work \citet{2023A&C....4400728L}, this process identified 12,501 FR-I sources from 946,366 FIRST images, with a mean confidence score of 0.84. Each FIRST image was retrieved from the FIRST cutout server\footnote{\url{https://third.ucllnl.org/cgi-bin/firstcutout}} using the component coordinates from the 17 December 2014 version of the catalog as the central position. All extracted images had a fixed angular size of 3.96 arcmin. Although the RGCMT model demonstrates absolute superiority in recognition speed, it cannot achieve 100\% accuracy. Therefore, this study employs a two-step selection process for FR-I samples detected by the RGCMT model: a ridgeline flux density distribution-based identification method followed by visual inspection.

\begin{figure*}
    \centering
    \includegraphics[scale=0.41]{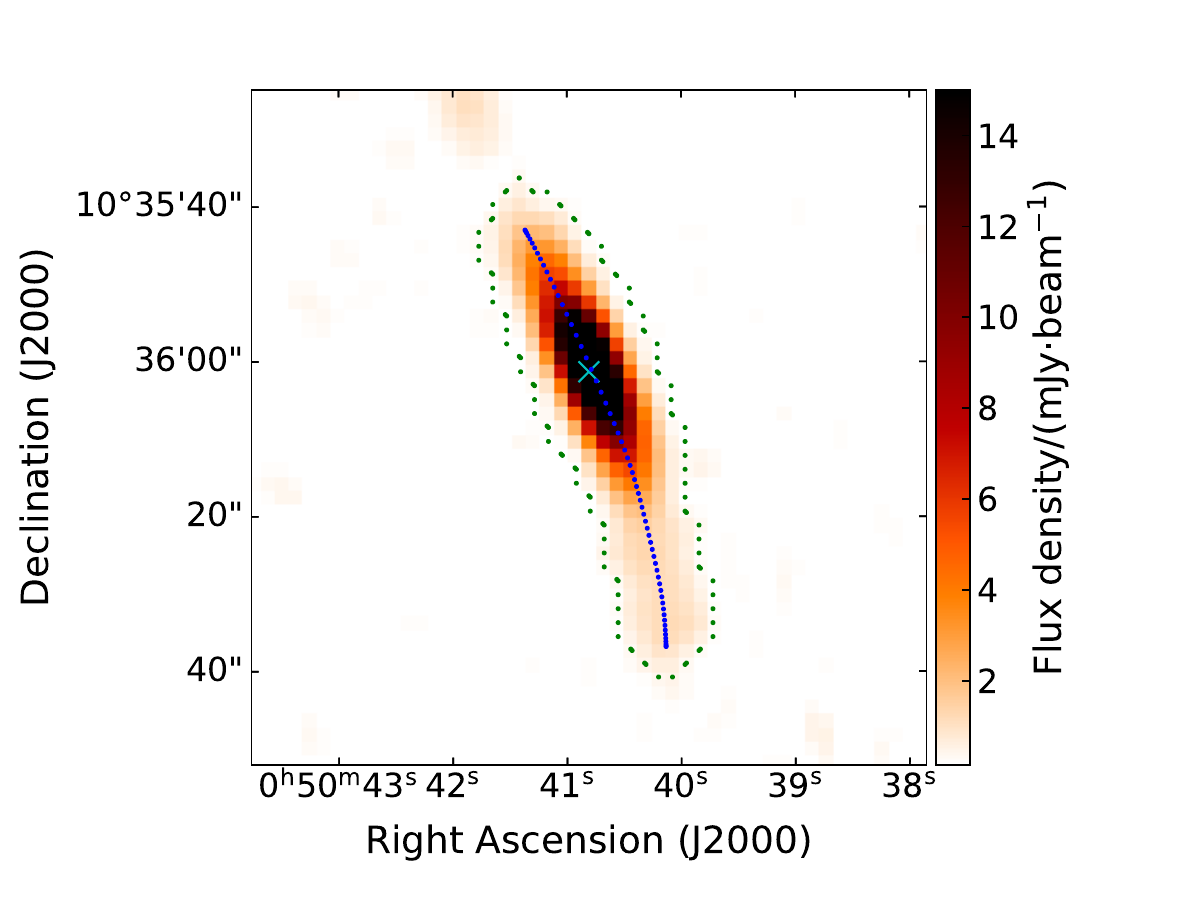}
    \hspace{2mm}
    \includegraphics[scale=0.41]{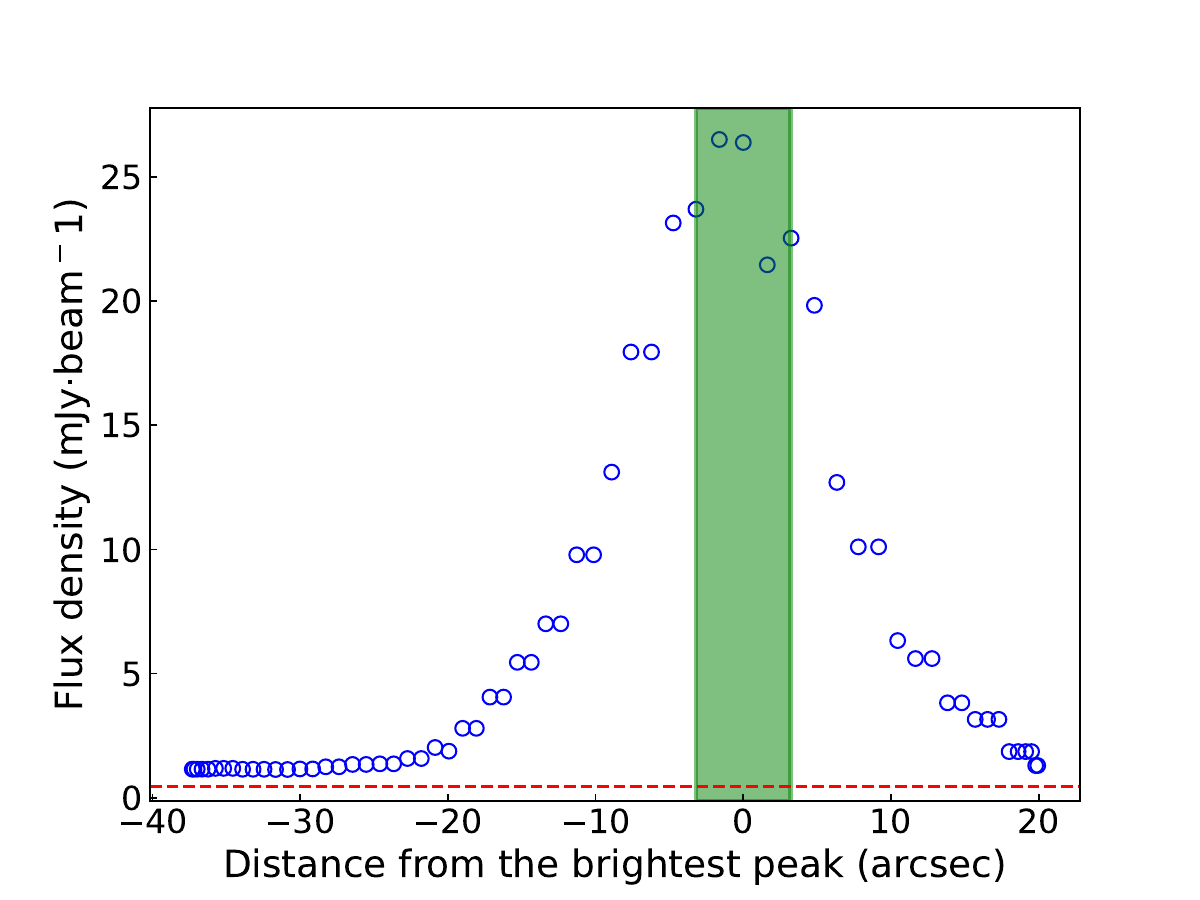}
    \includegraphics[scale=0.41]{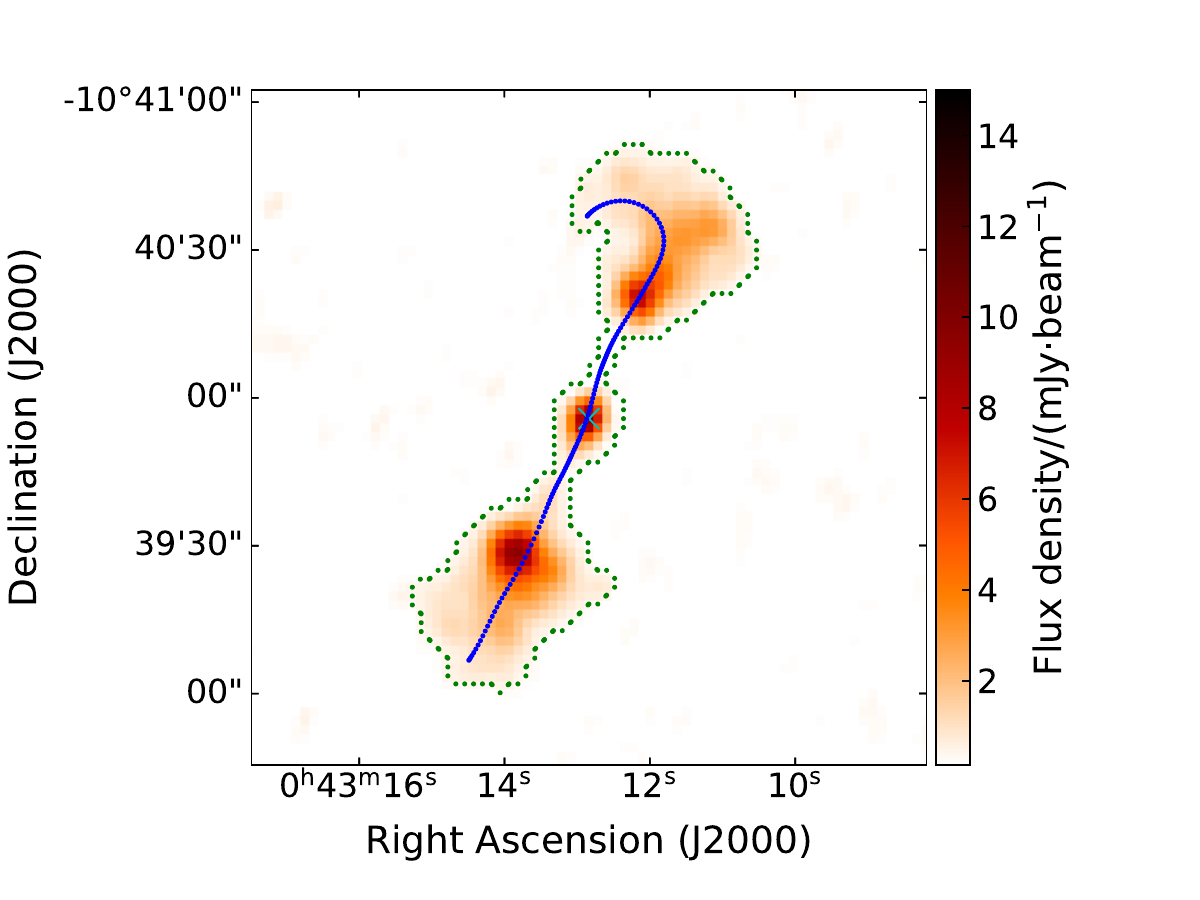}
    \hspace{2mm}
    \includegraphics[scale=0.41]{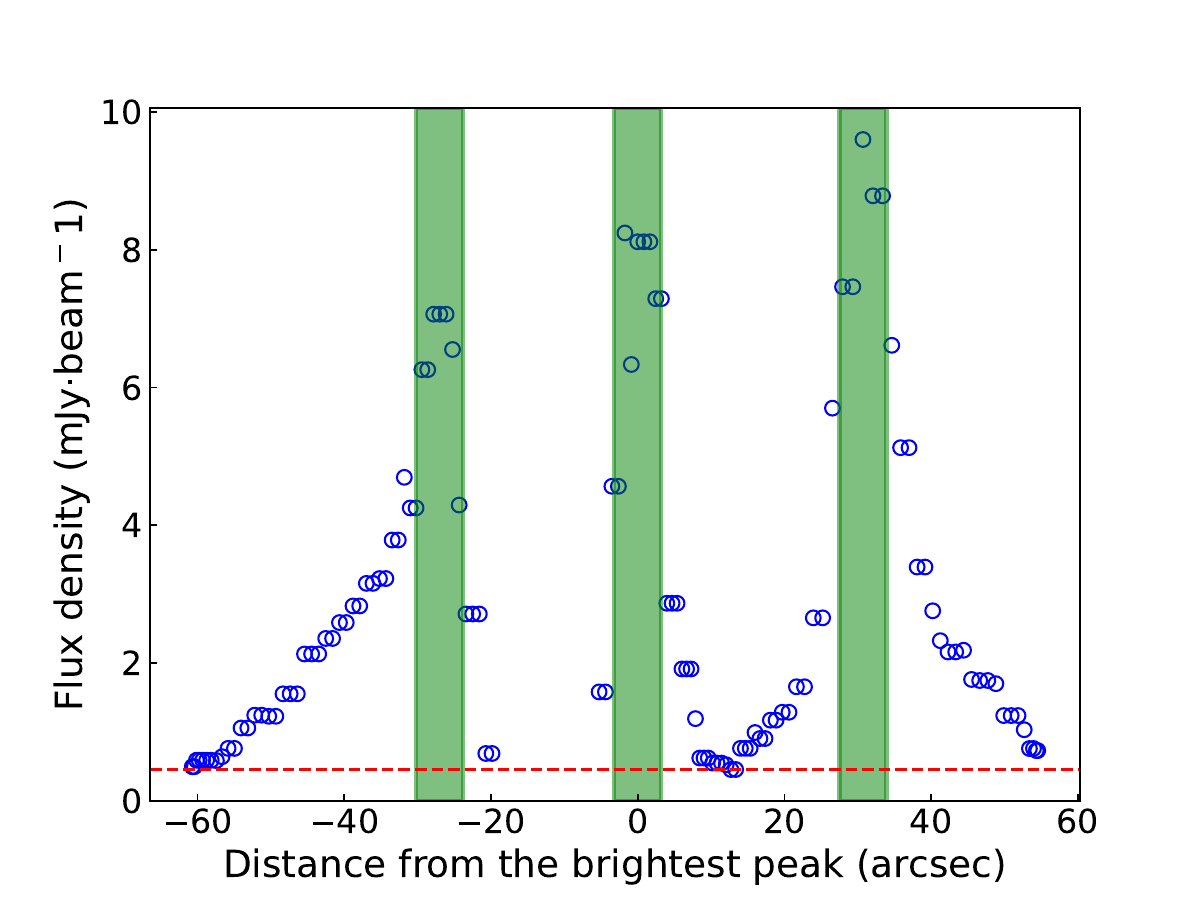}
    \includegraphics[scale=0.41]{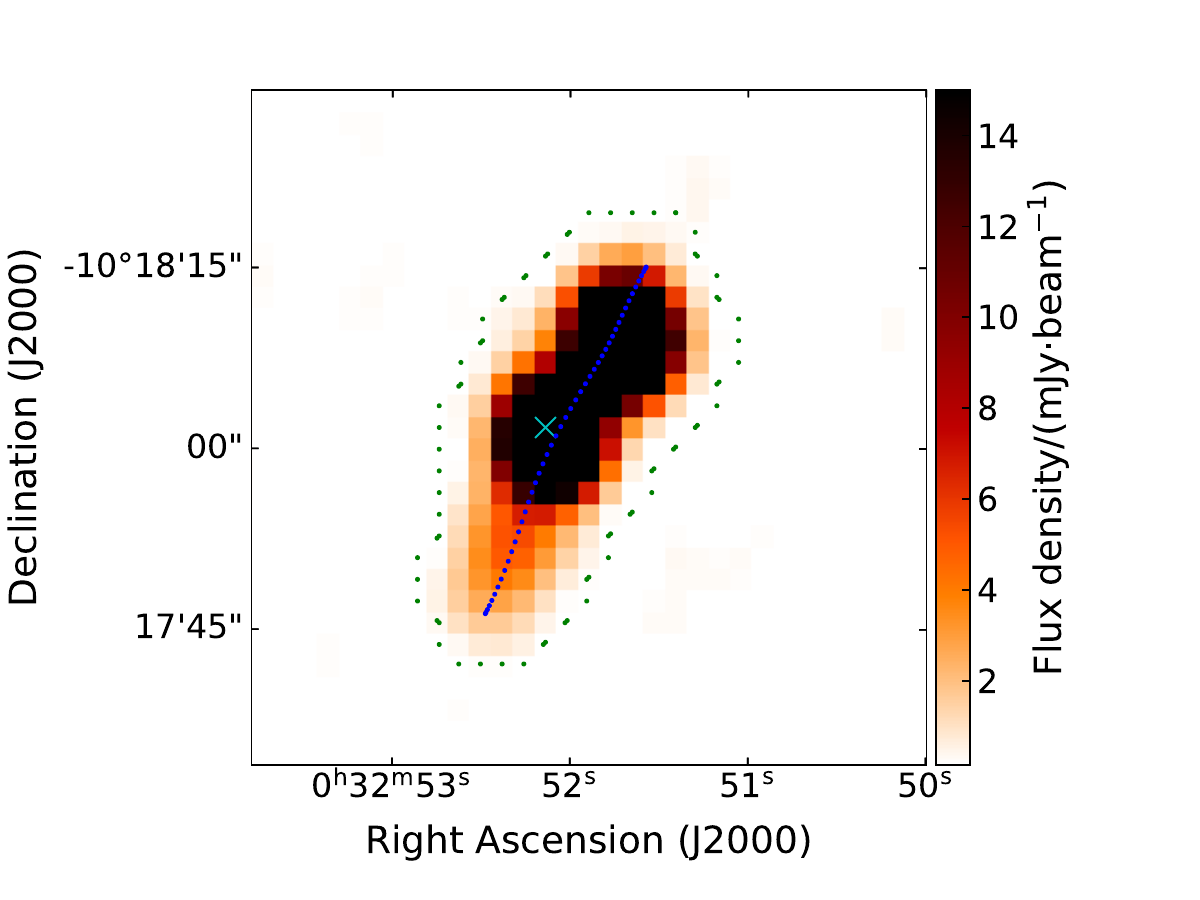}
    \hspace{2mm}
    \includegraphics[scale=0.42]{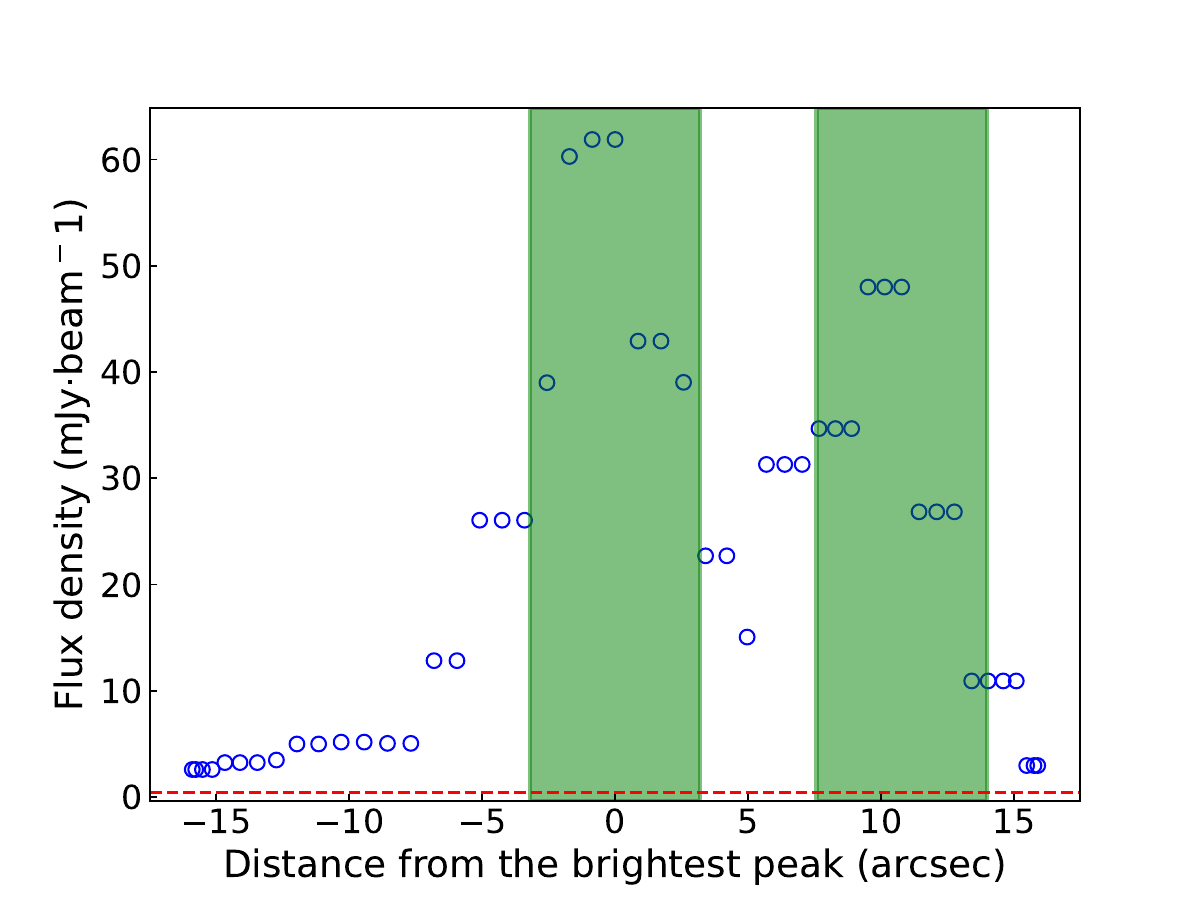}
    \caption{This figure shows how the ridgeline flux density distribution-based identification method applies to two typical morphological structures (upper and middle) in FR-I sources, as well as to one hybrid morphology radio source (bottom). Left panel: FR-I source from FIRST data with the source's boundary points (green) derived from the predicted mask, the peak flux density position (cyan $\times$), and the ridgeline (blue dotted line). Right panel: the flux density distribution at the ridgeline points. The x-axis represents the distance of each ridgeline point from the origin point, which is defined as the ridgeline point closest to the location of the brightest peak flux density. The green rectangular boxes are centered on the x-axis at positions corresponding to the peaks in flux density, with their widths equal to the local beam sizes of the FIRST image data. The red dashed line represents a flux density equal to 3 times the local RMS noise.
    \label{fig:ridgeline_example}}
\end{figure*}

Building upon the standard FR classification proposed by \citet{1974MNRAS.167P..31F}, subsequent studies have revised it, suggesting that FR morphology is generally more significant than the radio power break \citep[e.g.,][]{2017A&A...601A..81C,2019MNRAS.488.2701M}. A notable example is the work by \cite{2019MNRAS.488.2701M}, who proposed a novel automated morphological classification method for radio galaxies within the traditional FR framework. Their method classifies sources by comparing the distances to emission peaks against half of the total extent; however, this discrete criterion fails to account for spatial continuity and flux-density gradients along the jets. To address this limitation and ensure sample reliability, we developed a ridgeline-based approach using the mask predicted by the RGCMT model. This method traces the flux density distribution along each FR-I jet to quantify its monotonicity (i.e., the edge-darkened characteristic). In this method, we adhere to the standard FR-I morphological definition established by \citet{1974MNRAS.167P..31F}, which characterizes FR-I sources by their bright centers and faint edges. According to the ridgeline flux density distribution, an FR-I source must exhibit at least one central peak, with the flux density on both sides demonstrating a distinct monotonic decrease (continuing over a range equal to or larger than the local beam size) outward from this peak region (which may contain one or multiple peaks). Sources that show this distinct monotonic decrease on only one side are classified as either a Hybrid Morphology Radio Source \citep[HyMoRS;][]{2022MNRAS.514.4290K,2017AJ....154..253K} or a CJ source. In such cases, we classify the source as a CJ if it exhibits only a single peak; otherwise, it is identified as a HyMoRS.

An example of identifying FR-Is using the ridgeline-based method is shown in Figure \ref{fig:ridgeline_example}. The left panel presents the ridgeline of the source, which was measured in two steps: (1) the predicted mask of each source was transformed into polygon points (see the green points in the left panel), and (2) a Voronoi diagram \citep{aurenhammer1991voronoi} was generated from these points and then used to determine a ridgeline (blue dotted line) after applying a Gaussian smoothing filter. The right panel displays the ridgeline flux density distribution. The upper and middle panels display two characteristic FR-I morphological structures, featuring a single central peak and three distinct central peaks, respectively. In the upper panel, bilateral monotonic flux decay is observed beyond the central peak region (marked by the central green rectangle). Similarly, the middle panel exhibits pronounced monotonic decays to the left of the leftmost peak and to the right of the rightmost peak (indicated by the left and right green rectangles). The continuous decline range of the flux density on both sides of the peak regions exceeds 15 arcsec in both panels, a value that is significantly larger than the local beam size of 6.4 arcsec. In contrast, the bottom panel exhibits two distinct peaks. Beyond the peak regions, while the left side demonstrates a clear monotonic decrease, no significant monotonic decrease is observed on the right side, as its continuous descent range of 2 arcsec does not exceed the local beam size of 6.4 arcsec. This asymmetric flux profile confirms the source as a HyMoRS. All green rectangular boxes in the figure have x-axis dimensions matching the local beam size of the FIRST image in the left panel. After applying the aforementioned selection criteria, our final sample comprises 971 FR-Is with robust morphological classifications.

\subsection{Host galaxy cross-matching}
The Dark Energy Spectroscopic Instrument Legacy Surveys \citep[DESI LS;][]{2019AJ....157..168D} offer superior depth and photometric precision, which significantly enhance the reliability of optical counterpart identification for radio galaxies compared to shallower surveys like Sloan Digital Sky Survey \citep[SDSS;][]{2000AJ....120.1579Y,2022ApJS..259...35A} or the first part of the Panoramic Survey Telescope \& Rapid Response System \citep[Pan-STARRS1;][]{2020ApJS..251....7F}. The tenth data release of the DESI LS (DESI LS DR10) provides $griz$-band optical imaging, augmented by four infrared bands (W1-W4 at 3.4, 4.6, 12, and 22 $\mu m$, respectively) from the Wide-field Infrared Survey Explorer \citep[WISE;][]{2010AJ....140.1868W} and its Near-Earth Object mission extension (NEOWISE), covering over 20,000 deg$^2$ of the extragalactic sky. As DESI LS DR10 fully covers our sample, the host galaxies are identified using its optical photometry.

The host galaxies of FR-Is typically exhibit a spatial coincidence with their radio emission centroids, consistent with their characteristic jet-dominated morphology.
We performed an initial cross-match between our FR-I sample and the DESI LS DR10 using the X-Match Service at Astro Data Lab \citep{2014SPIE.9149E..1TF,2020A&C....3300411N}. A matching radius of 30 arcseconds, centered on the FR-I radio centroids, was adopted. This radius was chosen to be larger than the maximum projected separation of the host galaxies in our sample to ensure completeness. Subsequent filtering removed host galaxy candidates that had a signal-to-noise ratios (S/N) $<$ 5 in any of the $g$, $r$, $z$, W1 or W2 bands for each FR-I source, or that fell outside the predicted bounding box of the FR-I source. The host galaxies were then identified via visual inspection of the remaining candidates, following conventional procedures \citep[e.g.,][]{2015MNRAS.453.2326B}. The visual inspection was performed by overlaying the VLA FIRST radio contours and candidate positions on the DESI LS DR10 $r$-band images. In accordance with standard practice, any host candidates confirmed as stars must be excluded after the visual inspection, with details provided at the end of this section.

\begin{figure}[ht!]
\centering
\includegraphics[scale=0.38]{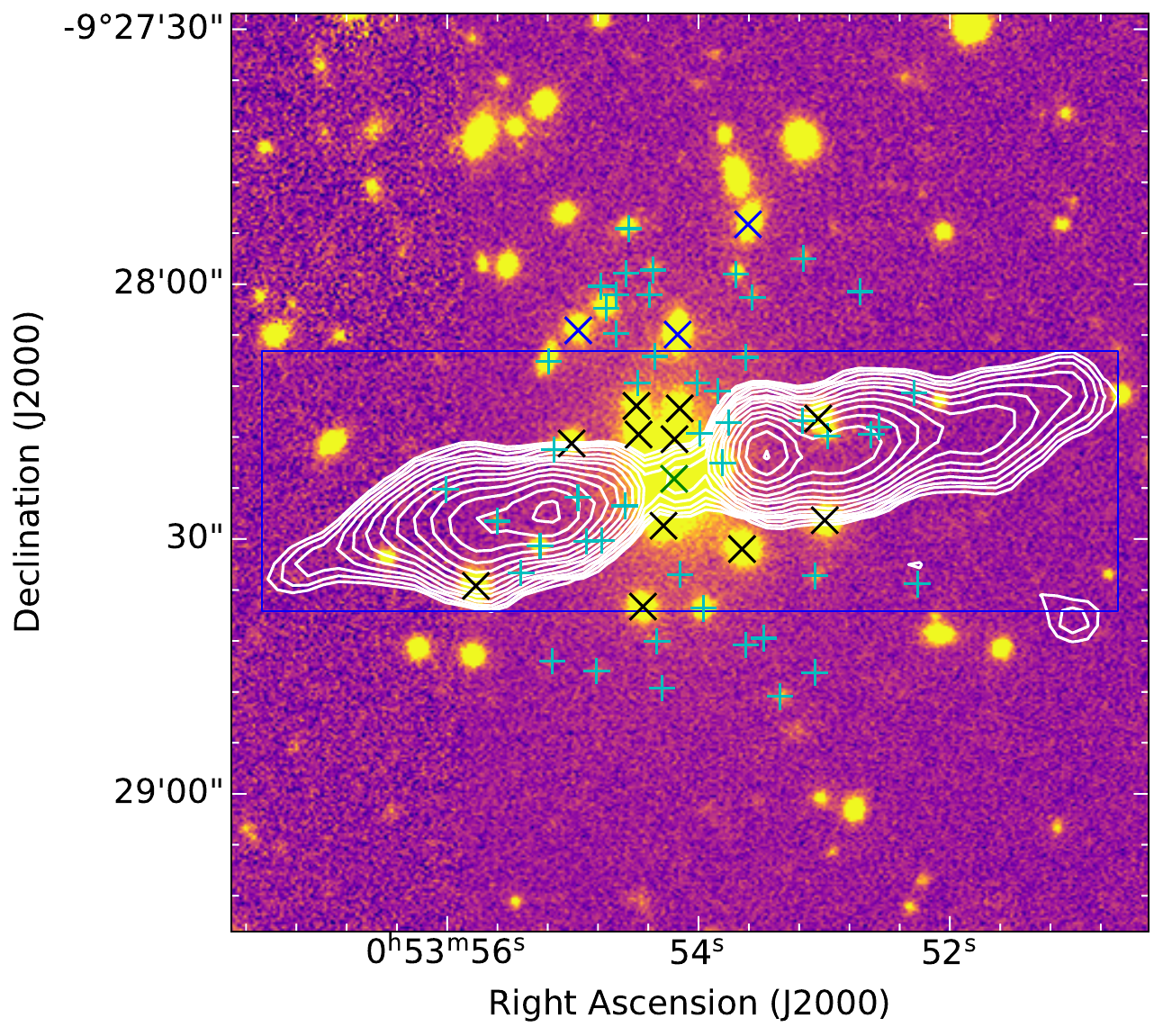}
    \caption{An example of the final host galaxy candidate identified for a FR-I radio galaxy from our sample. All matched host candidates are marked with `+' or `$\times$', where those with signal-to-noise ratio (S/N) $\geq$ 5 in any of the $g$, $r$, $z$, W1, or W2 bands are marked with `$\times$' and others with `+'. The final host candidate is marked with a green `$\times$'. The blue bounding box represents the predicted region of the FR-I source, while the blue `$\times$' markers indicate candidates located outside the predicted bounding box.
    \label{fig:example_hosts}}
\end{figure}

An example of host galaxy identification for an FR-I radio source is presented in Figure~\ref{fig:example_hosts}. A total of 62 host candidates were identified for this source. After applying the S/N cutoff, 47 candidate sources with S/N $<$ 5 were excluded, as indicated by the cyan `+' markers. Additionally, three high-S/N candidates, marked with blue `$\times$' symbols, were excluded for lying outside the estimated bounding box of the FR-I radio emission region. Visual inspection of the remaining candidates led to the removal of 11 further sources, leaving the candidate spatially coincident with the FR-I core as the final host candidate, marked by the green `$\times$' symbol. By applying this procedure, we successfully identified the final host candidates for 819 FR-Is.

\begin{figure*}[ht!]
\centering
    \includegraphics[scale=0.45]{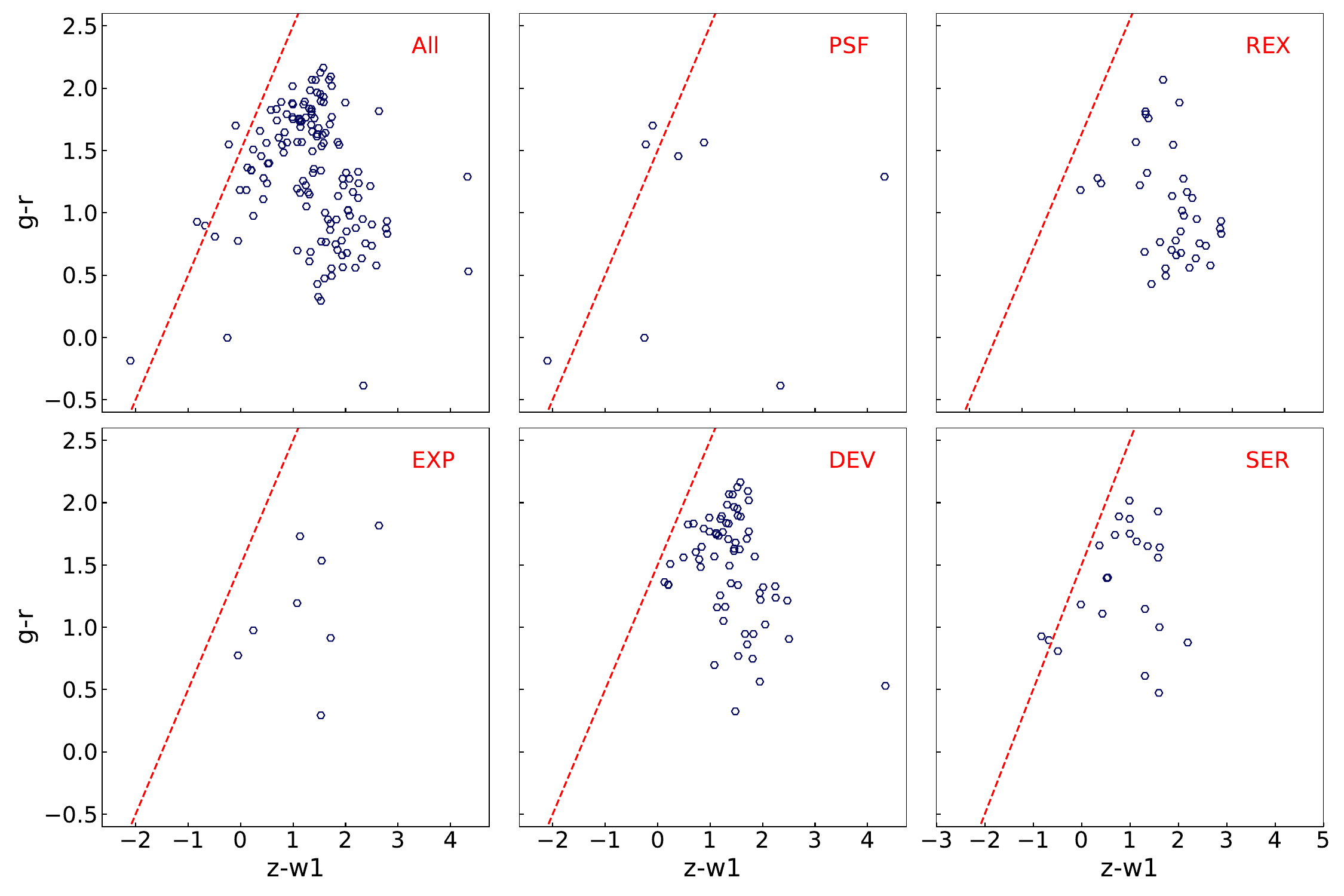}
    \caption{The $g-r$ vs. $z-$W1 color-color diagram for the 142 undetermined host candidates, classified by types. Labels PSF, REX, EXP, DEV, and SER correspond to host candidates with the following light-profile types: Point Spread Function, Round Exponential, deVaucouleurs, Exponential, and S\'{e}rsic, respectively. The label `ALL' designates the combined set of host candidates, incorporating all five morphological types.
    \label{fig:star-galaxy}}
\end{figure*}

In the final selection phase, all stellar sources were systematically excluded from the final host candidate list. We first cross-matched the final host candidate list with both the first data release of the DESI \citep[DESI DR1;][]{2025arXiv250314745D} and the Seventeenth Data Release of the SDSS \citep[SDSS DR17;][]{2022ApJS..259...35A} using a matching radius of 1 arcsec. The result produced 570 matched host candidates: 263 from DESI DR1 and 307 from SDSS DR17. One candidate was then excluded from the final host candidate list based on its spectroscopic classification as a star in SDSS DR17. For the remaining host candidates without counterparts in either DESI DR1 or SDSS DR17, we performed cross-matching with the galaxy catalog from DESI LS Data Release 8 \citep{2023MNRAS.526.4768W} and the quasi-stellar object (QSO) catalog from DESI LS Data Release 9 \citep{2022RAA....22i5021H}, adopting a 1 arcsec search radius to classify their nature. Through cross-matching, an additional 90 host candidates were classified as galaxies and 17 were identified as QSOs. For the remaining 142 undetermined host candidates, we applied the star-galaxy separation method from Astro Data Lab Jupyter notebook \citep{Astrodatalabnotebook} to classify them, using the optical colour $g-r$, the near-IR to mid-IR colour $z-$W1, and the object type classification in the DESI LS DR10. In the DESI LS DR10, the sources are classified into five object types: Point Spread Function (PSF; point sources), Round Exponential profile (REX; round exponential galaxies), deVaucouleurs profile (DEV; elliptical galaxies), Exponential profile (EXP; spiral galaxies), and S\'{e}rsic profile (SER). The label `ALL' denotes the set of host candidates encompassing all types. Typically, PSF objects are mostly stars but may include some QSOs, while the vast majority of extended objects classified as REX, EXP, DEV, or SER are galaxies. The $g-r$ vs. $z-$W1 color-color diagram for the remaining 142 undetermined host candidates (see Figure~\ref{fig:star-galaxy}), categorized by object type. The star-galaxy separation is marked by a red dashed line following ($g-r$)=($z-$W1)+1.5. Candidates above the red dashed line are classified as stars, while those below are either galaxies or QSOs. In this study, we classify host candidates located below the red dashed line according to their types: sources exhibiting PSF profiles are identified as QSO candidates (Qc), while those showing non-PSF morphology are classified as galaxy candidates (Gc). Based on this classification scheme, we identify 5 stars that are subsequently removed from the final host candidate list, leaving 132 Gc and 5 Qc. We therefore successfully identified the true host galaxies for 813 FR-I sources in our sample.

\subsection{Spectroscopic and photometric redshifts}
Redshift serves as the fundamental observable property of radio galaxies, providing the essential cosmological distance metric required for deriving key physical characteristics, including radio luminosity and largest projected linear size (LLS). Spectroscopic redshifts are generally more reliable than photometric estimates, as they are derived from the precise measurement of atomic/molecular emission or absorption lines, rather than broadband spectral energy distribution (SED) fitting. We cross-matched our FR-I hosts with major spectroscopic redshift catalogs to obtain reliable redshifts, including DESI DR1, SDSS DR17, the 2dF Galaxy Redshift Survey \citep[2dFGRS;][]{2001MNRAS.328.1039C}, the 6dF Galaxy Survey \citep[6dFGS;][]{2009MNRAS.399..683J}, the WiggleZ Dark Energy Survey \citep{2018MNRAS.474.4151D}, the Two Micron All Sky Survey \citep[2MASS;][]{2006AJ....131.1163S} Redshift Survey \citep[2MRS;][]{2012ApJS..199...26H}, and the NASA/IPAC Extragalactic Database (NED)\footnote{\url{https://ned.ipac.caltech.edu}}. A conservative matching radius of 1 arcsec was adopted to minimize spurious associations while ensuring robust counterpart identification. For sources with multiple redshift measurements, we prioritized values from higher-resolution surveys (e.g., DESI DR1 or SDSS DR17). For the remaining objects without spectroscopic redshifts, we supplemented them with photometric redshifts (z\_phot\_mean) obtained from the photo\_z table of DESI LS DR10 by matching with the unique LS object ID (ls\_id) of their FR-I host galaxies. From a total sample of 813 FR-I hosts, we obtained spectroscopic redshifts for 628 (77.2\%). The breakdown by survey is as follows: 263 from DESI DR1, 305 from SDSS DR17, 1 from 2dFGRS, 17 from 6dFGS, 32 from 2MRS, 1 from the WiggleZ Dark Energy Survey, and 9 from NED. Photometric redshifts were successfully obtained for the remaining 185 FR-I host galaxies from DESI LS DR10.

\section{Results and discussion} \label{Sec:RE}

\subsection{The FR-I radio galaxies catalog}
\begin{figure*}
    \centering
	\includegraphics[scale=0.25]{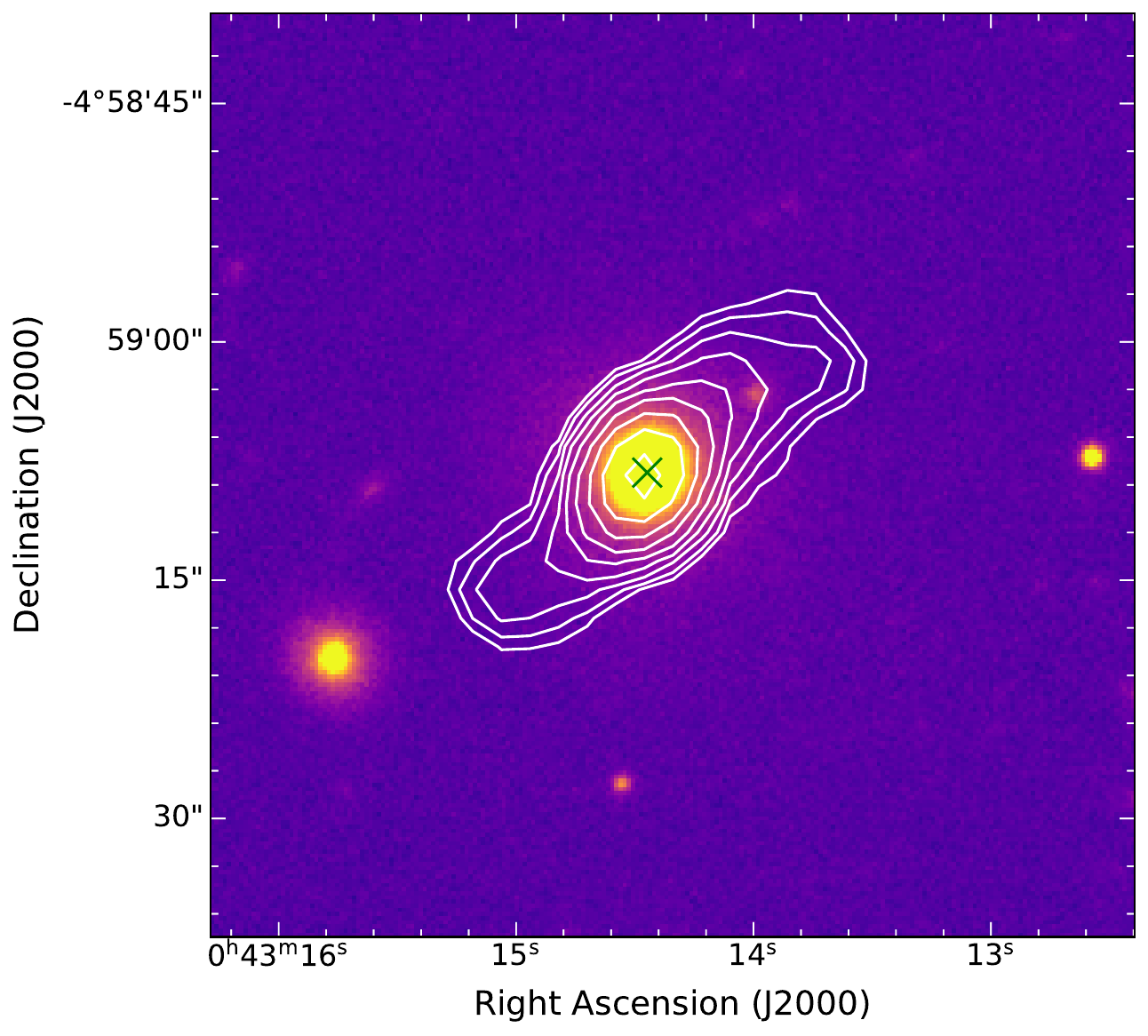} 
    \hspace{2mm}
    \includegraphics[scale=0.25]{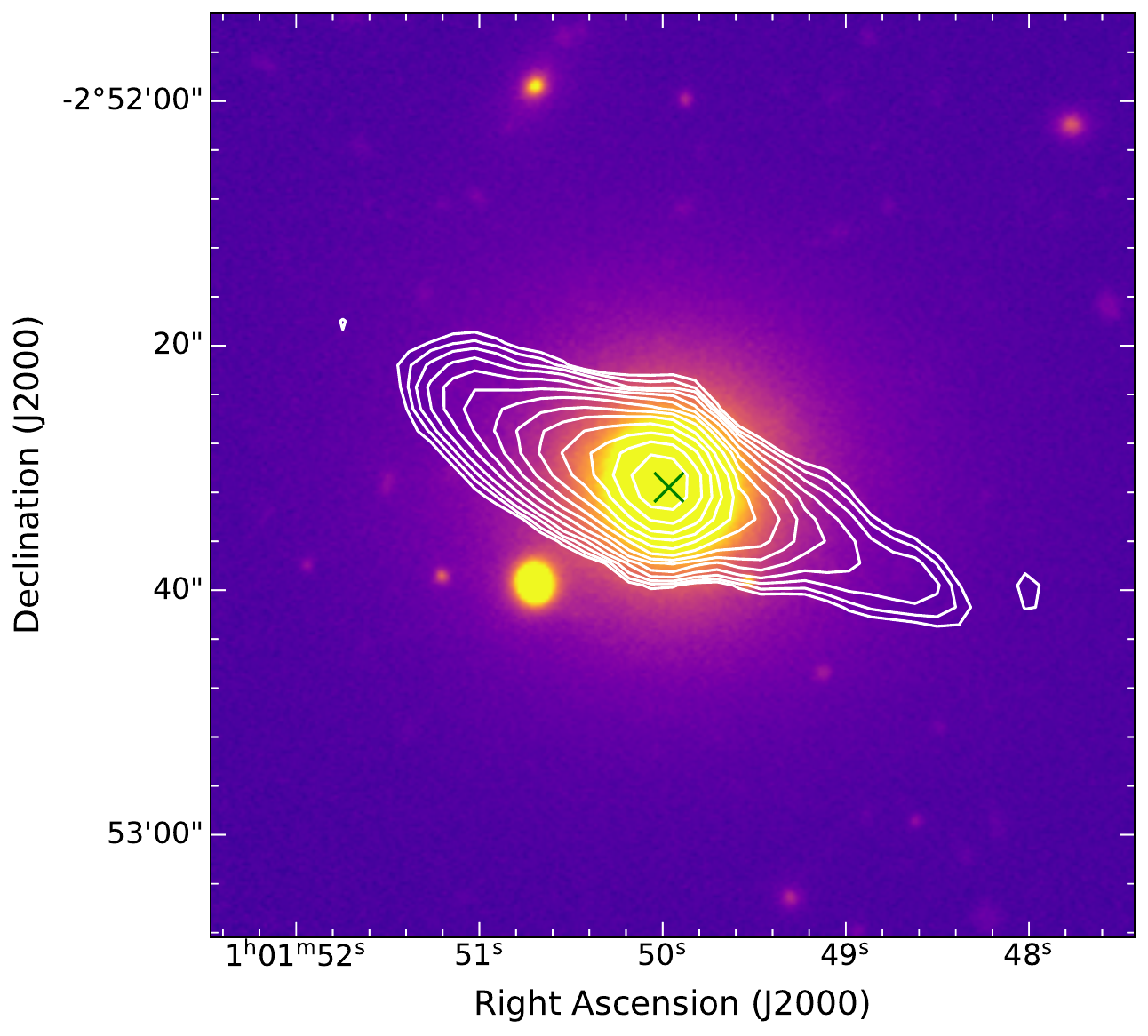} 
    \hspace{2mm}
    \includegraphics[scale=0.25]{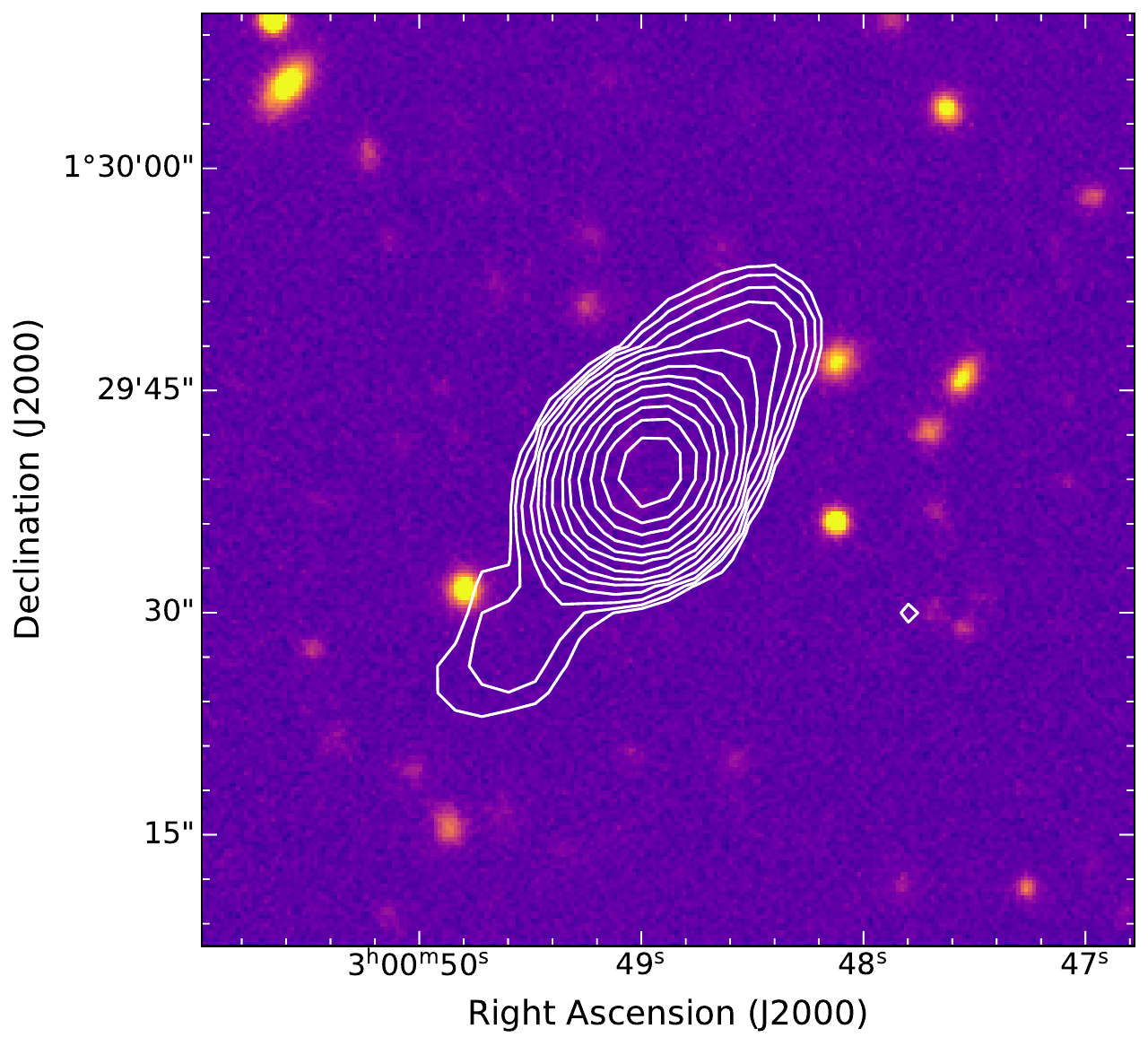} 
    \hspace{20mm}
    \includegraphics[scale=0.25]{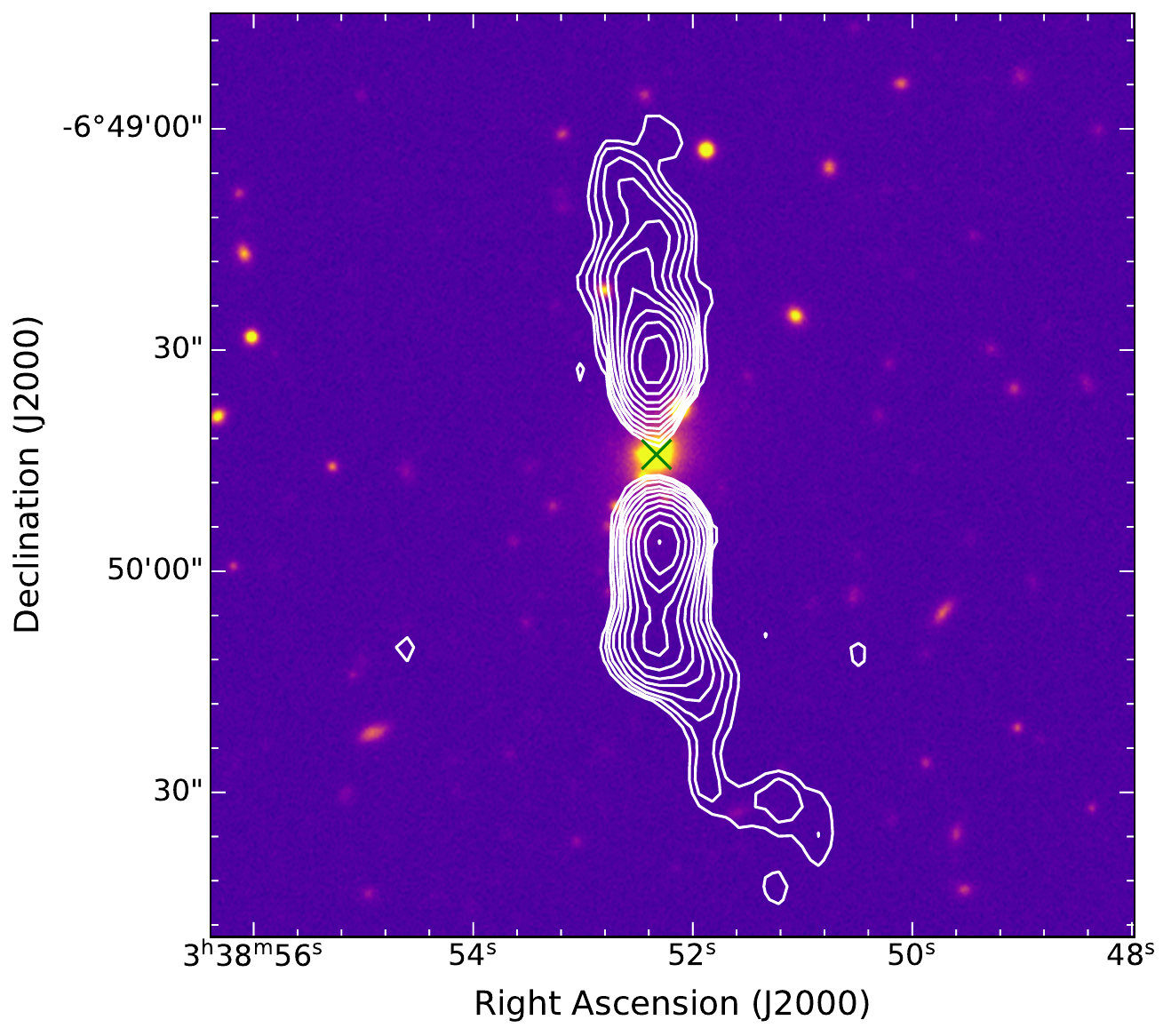} 
    \hspace{2mm}
    \includegraphics[scale=0.25]{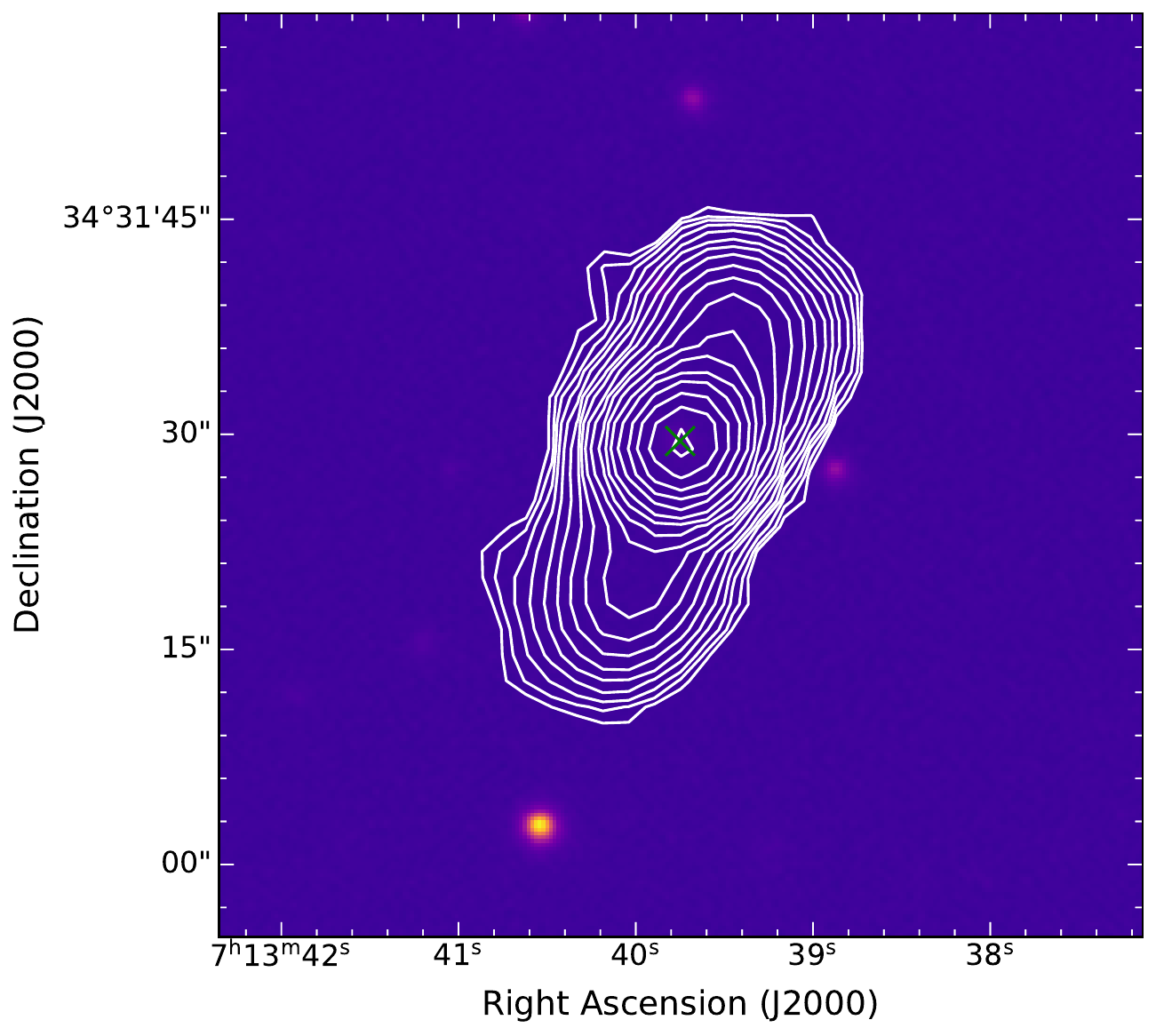} 
    \hspace{2mm}
    \includegraphics[scale=0.25]{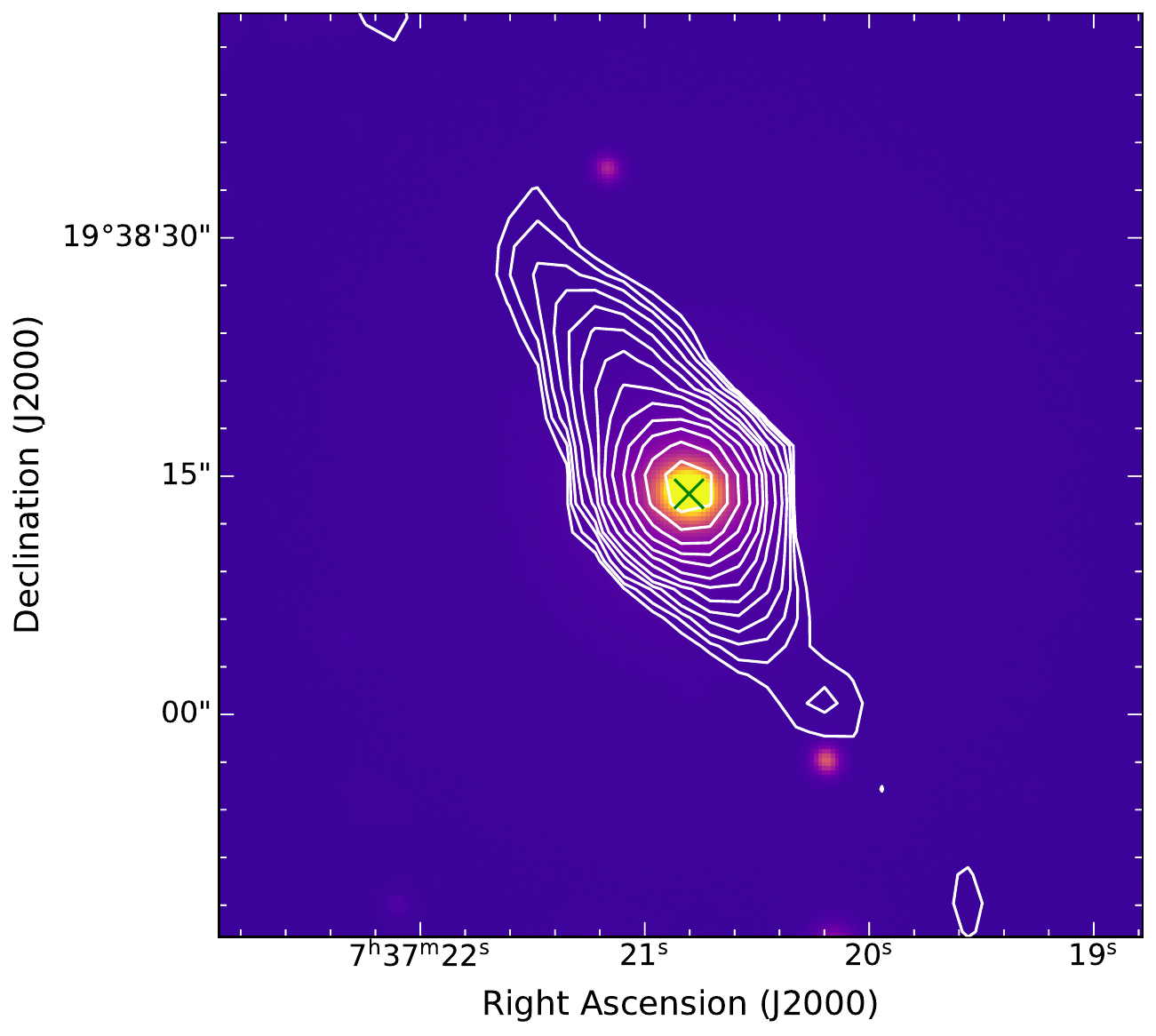} 
    \hspace{20mm}
    \includegraphics[scale=0.25]{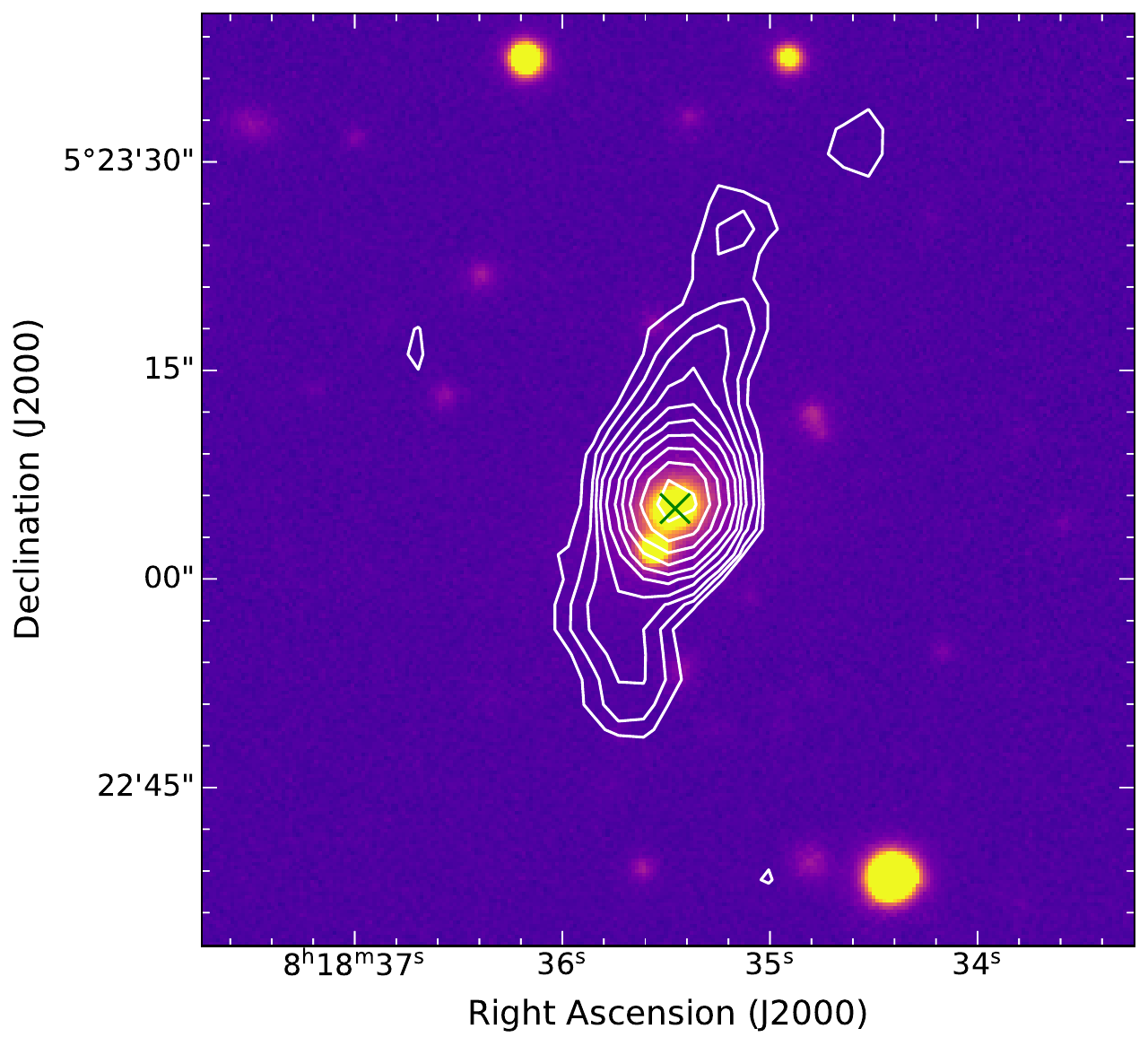}
    \hspace{2mm}
    \includegraphics[scale=0.25]{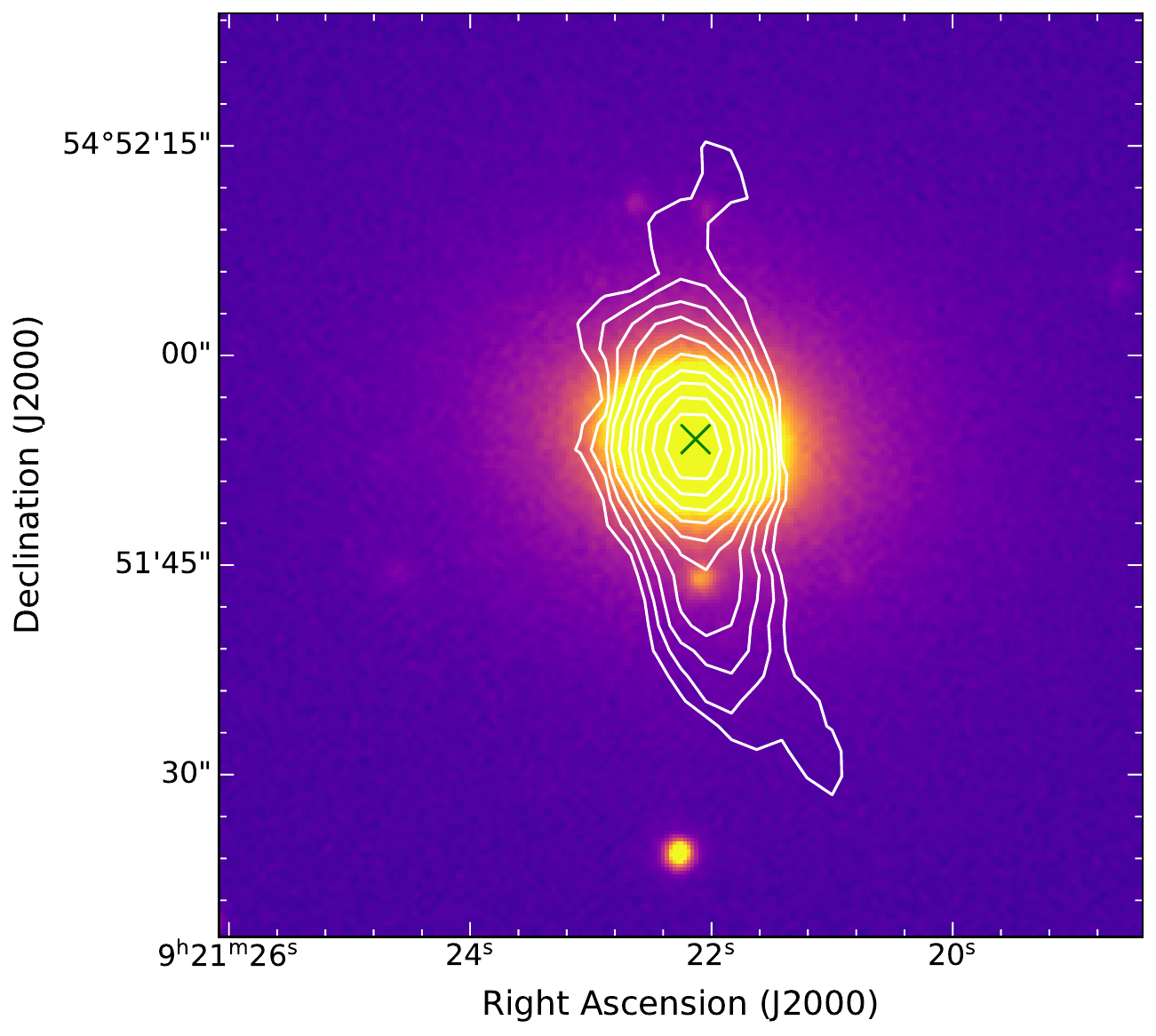} 
    \hspace{2mm}
    \includegraphics[scale=0.25]{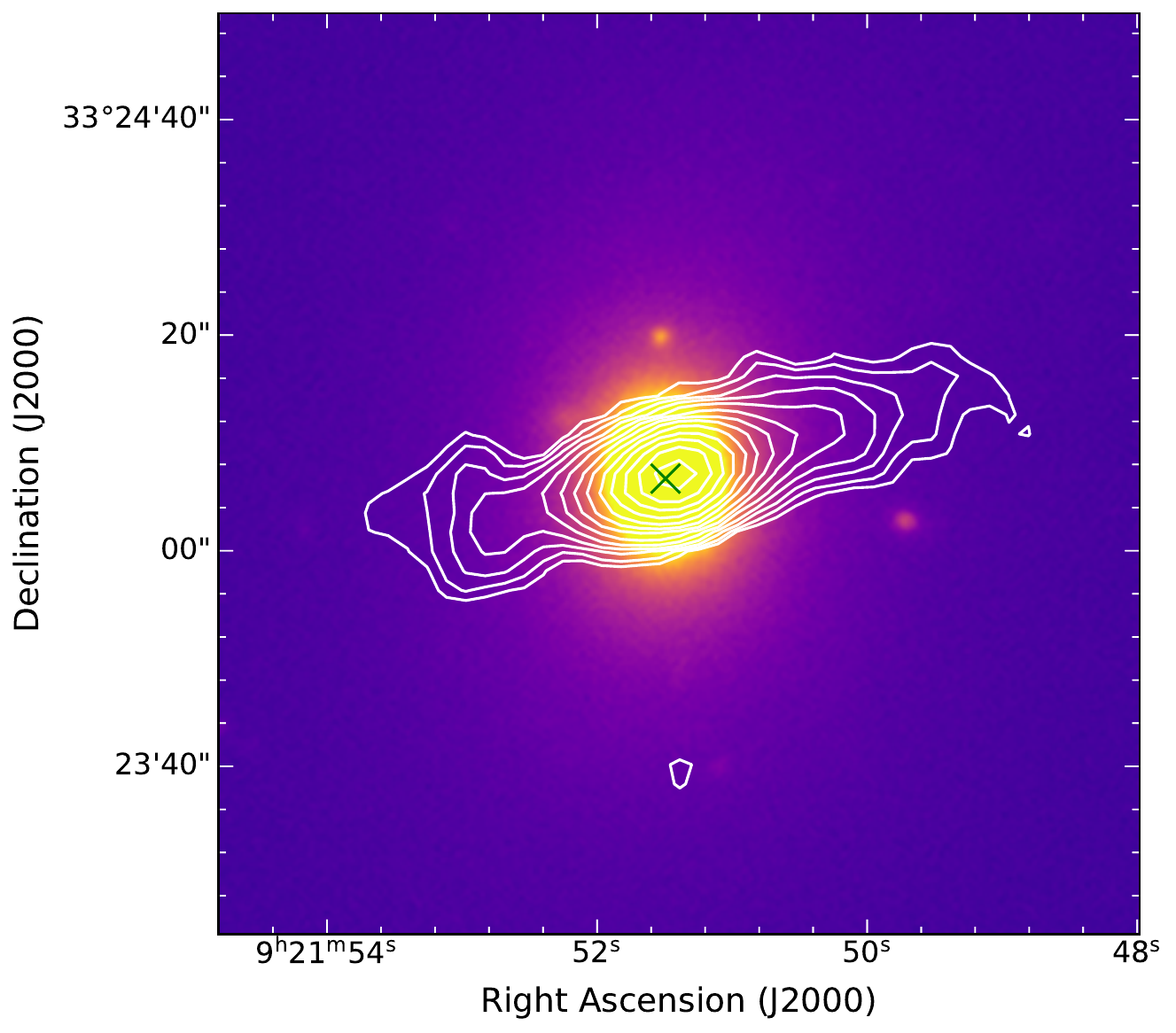} 
    \hspace{20mm}
    \includegraphics[scale=0.25]{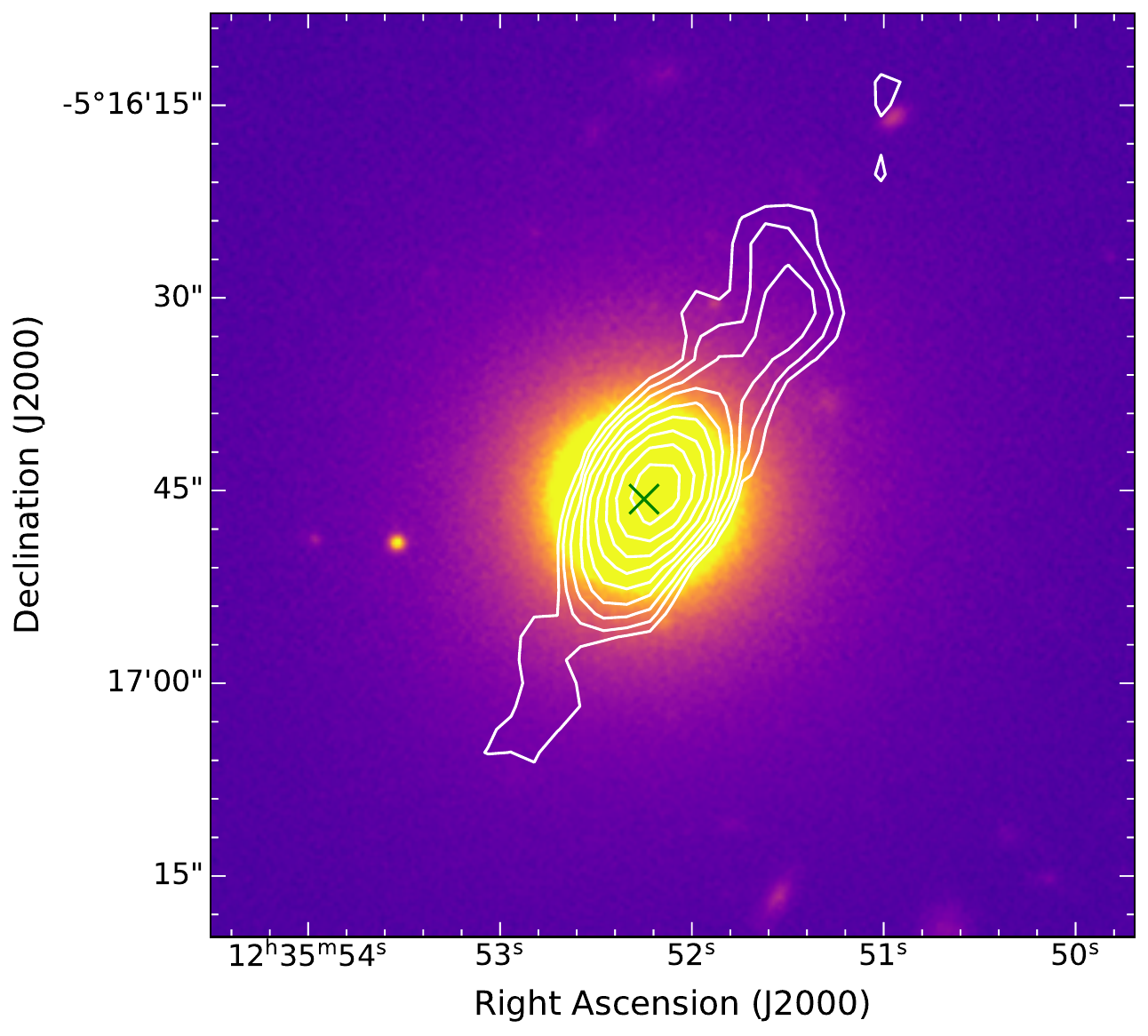} 
    \hspace{2mm}
    \includegraphics[scale=0.25]{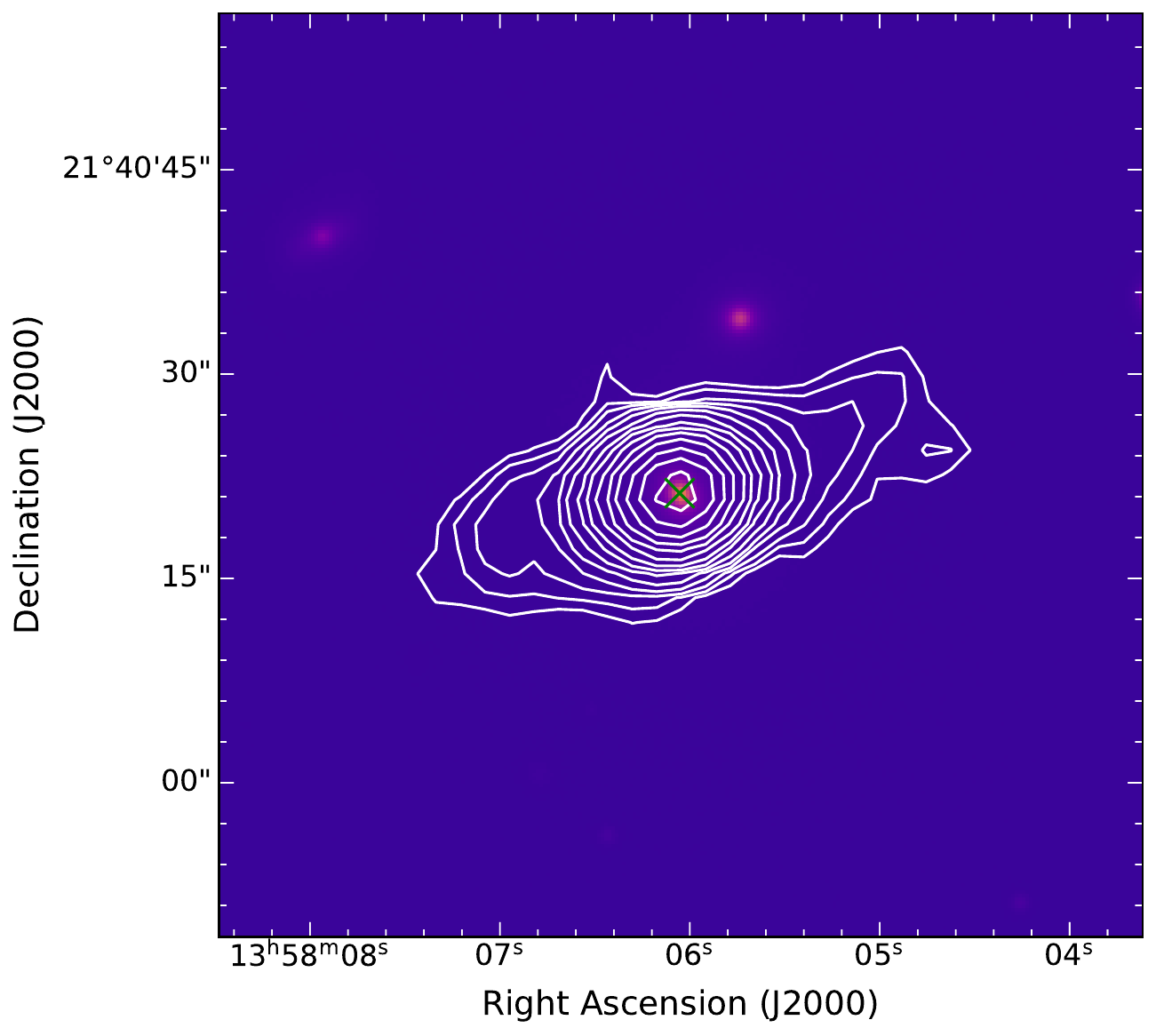}
    \hspace{2mm}
    \includegraphics[scale=0.25]{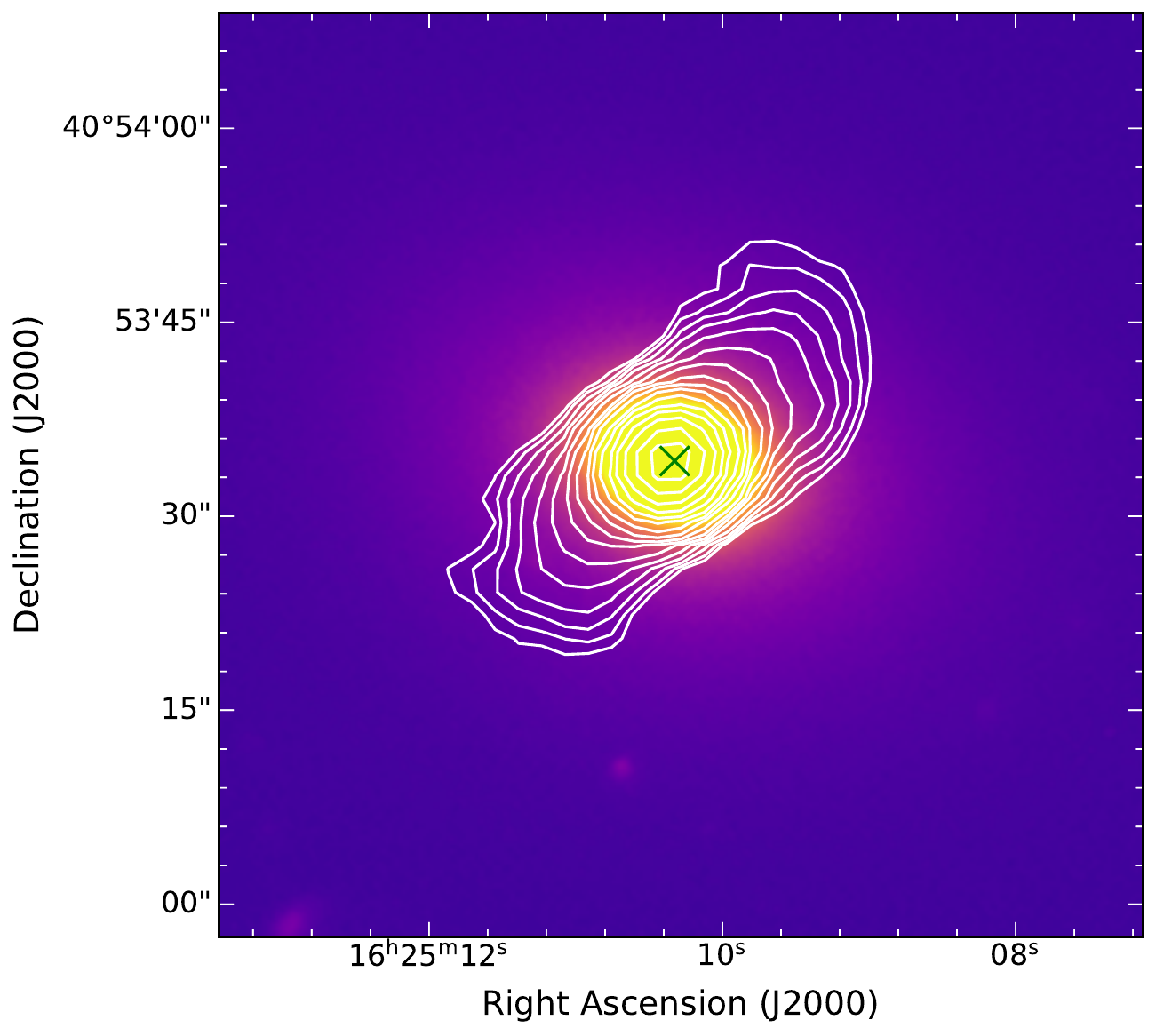} 
    \caption{Radio-optical overlay of 12 newly identified FR-Is from our catalog, showing VLA FIRST radio continuum contours overlaid on DESI LS $r$-band optical images (plasma colour scale). The contour levels start at 3$\sigma_{\rm rms}$ (where $\sigma_{\rm rms}$ is the local RMS noise) and increase by multiplicative factors of $\sqrt{2}$. The host galaxy positions are indicated by green $\times$ symbols.
    \label{fig:fr1_example}}
\end{figure*}

We present a catalog of FR-I radio galaxies selected from the VLA FIRST survey, as shown in Table~\ref{tab:radio_catalog}. Columns (1) and (2) list the short names and full names, respectively. For FR-I sources with identified host galaxies, we adopt the host galaxy names in column (2); otherwise, the names are derived from the centroid coordinates in the RGCMT model. The host galaxy names follow the redshift-reference catalog or imaging survey from which the spectroscopic or photometric redshift was obtained. The source name flags in column (2) denote cross-identification with previous catalogs: `a’ for sources from \citet{2017AA...598A..49C}, `b’ for those in \citet{2017MNRAS.466.4346M}, `c’ for entries from \citet{2022MNRAS.509.2269S}, and `d’ for objects listed in \citet{2023DIB....4708974G}. Our FR-I sample contains 53, 46, 83, and 75 sources flagged with `a', `b', `c', and `d', respectively. There is significant overlap among these four catalogs. After de-duplication, we found that 126 FR-Is were previously cataloged. The incomplete overlap between our catalog and the four reference FR-I catalogs primarily results from our distinct classification scheme. Our sample explicitly excludes morphologically bent FR-I sources, such as J103827.01$+$414852.9, J122951.84$+$114024.2, and J125434.92$-$023412.4 found in the reference catalogs, as well as one-sided extended or CJ sources including J004300.63$-$091346.3, J002900.98$-$011341.7, and J015253.79$-$001005.5 from these catalogs. Figure~\ref{fig:fr1_example} shows the radio-optical overlays of 12 examples of newly discovered FR-Is.

\begin{deluxetable*}{lccccccccccc}
\digitalasset
\tablewidth{0pt}
\setlength{\tabcolsep}{5pt}
\tablecaption{A catalog of FR-I radio galaxies from the VLA FIRST survey. \label{tab:radio_catalog}}
\tablehead{
   \colhead{Short name} & \colhead{Name} & \colhead{R.A.}  & \colhead{Decl.} & \colhead{Typ.} & \colhead{$z$} & \colhead{$F_{1400}$} & \colhead{$F_{150}$} & \colhead{$\alpha_{150}^{14000}$} & \colhead{LAS} & \colhead{LLS}  & \colhead{${\rm log}_{10}(P_{1400})$} \\
    \colhead{(J2000)} & \colhead{(J2000)} & \colhead{(deg)}  & \colhead{(deg)} & \colhead{} & \colhead{} & \colhead{(mJy)} & \colhead{(mJy)} & \colhead{($\pm$0.05)} & \colhead{(arcsec)} & \colhead{(kpc)} & \colhead{(${\rm{W}}\,{{\rm Hz}^{ - 1}}$)} 
}
\colnumbers
\startdata
J0001+1250 &SDSS J000108.01+125003.6 &0.28338 &12.83434 & G &0.200$^{\rm S}$ &14.2$\pm$0.7 & 42.0$\pm$4.1 &-0.48 &32 &105 &24.17 \\
J0003+0709 &DESI LS J000313.52+070945.3 &0.80634 &7.16257 &G &0.208$^{\rm P}$ &48.0$\pm$2.3 &135.9$\pm$12.8 &-0.47 &29 &98 &24.74 \\
J0003+0134 &DESI LS J000337.82+013402.7 &0.90758 &1.56742 &Gc &0.966$^{\rm P}$ &110.4$\pm$5.6 &37.4$\pm$3.4 &0.48 &36 &285 &26.29 \\
J0009+1237 &RGCMT J000941.69+123725.6 &2.4237 &12.62378 & -- & --&80.0$\pm$3.8 &23.4$\pm$2.2 & 0.55 & 73 & --& -- \\
J0010-0625 &RGCMT J001034.47-062539.6 &2.64362 &-6.42767 &--&-- &14.4$\pm$0.7 &95.8$\pm$8.9 &-0.85 &24 &--& -- \\ 
J0012-0723 &SDSS J001211.46-072344.1 &3.04774 &-7.39558 &G &0.319$^{\rm S}$ &29.9$\pm$1.5 &80.0$\pm$7.4 &-0.44 &31 &144 &24.93 \\
J0012-0857 &SDSS J001249.21-085733.1 &3.20503 &-8.9592 &G &0.335$^{\rm S}$ &58.0$\pm$2.9 &195.6$\pm$18.9 &-0.54 &34 &163 &25.28 \\
J0015-0653 &SDSS J001503.66-065306.8 &3.76525 &-6.88522 &G &0.622$^{\rm S}$ &61.3$\pm$3.0 &759.5$\pm$70.2 &-1.13 &46 &312 &26.03\\
J0015+1440 &RGCMT J001532.07+144014.6 &3.88363&14.67071&--&--&203.2$\pm$10.0 &750.7$\pm$70.1 &-0.59 &63 &--& --\\
J0017-0512 &DESI J001735.82-051241.8 &4.39924 &-5.2116 &Q &0.227$^{\rm S}$ &193.3$\pm$9.4 &316.7$\pm$29.9 &-0.22 &41 &149 &25.4\\
J0018+1117 &DESI J001805.04+111751.2 &4.52102 &11.29755&G &0.136$^{\rm S}$ &81.1$\pm$4.1&301.2$\pm$27.5 &-0.59 &52 &125 &24.58 \\
J0019+1155 &DESI J001940.52+115556.1 &4.91881 &11.93226&G &0.449$^{\rm S}$ &97.9$\pm$4.8 &372.6$\pm$35.5 &-0.6 &41 &235 &25.80 \\
J0026-0556 &DESI LS J002607.76-055602.2 &6.53235 &-5.93394 &Q &1.345$^{\rm P}$ &24.9$\pm$1.2 &87.9$\pm$8.8 &-0.56 &34 &285 &26.27 \\
J0027+0947 &RGCMT J002738.37+094723.2 &6.90987 &9.78977 &--&--&15.2$\pm$0.7 &30.1$\pm$3.0 &-0.31 &38 &--& --\\
J0027-0204 &DESI LS J002757.24-020414.9 &6.98849 &-2.07079&G &0.225$^{\rm P}$ &73.9$\pm$3.6 &224.8$\pm$21.8 &-0.5 &30 &108 &25.00\\
J0031+1212 &SDSS J003133.41+121227.0 &7.88923 &12.20749&G &0.177$^{\rm S}$ &65.4$\pm$3.2 &135.0$\pm$13.3 &-0.32 &39 &116 &24.71 \\
J0034-0638 &DESI LS J003434.10-063857.4 &8.64207 &-6.64929 &Gc &0.949$^{\rm P}$ &27.7$\pm$1.4 &105.4$\pm$9.9 &-0.6 &32 &252 &25.99 \\
J0038-0207 &6dF J0038205-020741 &9.58546 &-2.12792 &G &0.221$^{\rm S}$ &6206.0$\pm$315.4 &29955.0$\pm$2797.5 &-0.7 &66 &235 &26.92 \\
J0039-1032$^{\rm a,d}$ &SDSS J003930.53-103218.6 &9.8772 &-10.53851 &G &0.129$^{\rm S}$ &31.6$\pm$1.5 &81.1$\pm$7.4 &-0.42 &30 &69 &24.11\\
J0039+1108	&RGCMT J003942.06+110852.0	& 9.92523	&11.14777 &-- &--	& 18.8$\pm$0.9 &	73.8$\pm$7.3 &	-0.61	& 28	& --& --\\ 	
\enddata
\tablecomments{Column (1) short name (JHHMM+DDMM); flags `a', `b', `c', and `d' in the short names indicate the sources present in \citet{2017AA...598A..49C}, \citet{2017MNRAS.466.4346M}, \citet{2022MNRAS.509.2269S}, and \citet{2023DIB....4708974G}, respectively. Column (2): full name derived either from the host galaxy name (following the redshift-reference catalog or imaging survey) or from the centroid coordinates in the RGCMT model in J2000. Columns (3) and (4): coordinates corresponding to the full names in deg. Column (5): spectral type of the host. Column (6): redshift ($z$) with a flag indicating whether it is spectroscopic (`S') or photometric (`P'). Column (7): integrated flux densities at 1400~MHz in mJy. Column (8): integrated flux densities at 150 MHz in mJy. Column (9): spectral index between 150 MHz and 1400 MHz. Column (10): the largest angular size (LAS) in arcsec. Column (11): the largest projected linear size (LLS) in kpc. Column (12): the logarithm radio power in ${\rm{W}}\,{{\rm Hz}^{ - 1}}$ at 1400 MHz. The full version of Table \ref{tab:radio_catalog} is available at: \url{https://drive.google.com/file/d/1CHRAMRj40BkrQo7mfRysiTlLTJ5TfWyN/view?usp=sharing}.}
\end{deluxetable*}

The celestial coordinates in columns (3) and (4) provide the J2000 epoch Right Ascension (R.A.) and Declination (Dec.) for each source in decimal degrees. Column (5) specifies the host galaxy classification: `G' for confirmed galaxies, `Gc' for galaxy candidates, `Q' for confirmed QSOs, and `Qc' for quasar candidates. Column (6) provides the redshift, with a flag denoting whether it is a spectroscopic (`S') or photometric redshift (`P'). The redshifts of our FR-I sources span a range of 0.002$\leq  z \leq$2.307. Columns (7) and (8) provide the integrated flux densities at 1400 MHz and 150 MHz, respectively. The 1400 MHz flux densities were derived from the NRAO VLA Sky Survey \citep[NVSS;][]{1998AJ....115.1693C}, while the 150 MHz values were computed using data from the TIFR GMRT Sky Survey Alternative Data Release 1 \citep[TGSS ADR1;][]{2017A&A...598A..78I}. The 1400 MHz integrated flux densities of our FR-I sample range from 10.2 mJy to 6.2 Jy. The two-point spectral indexes between 150 and 1400 MHz ($\alpha_{150}^{1400}$) for the FR-I sample, listed in column (9), were calculated using the power-law relation $F_\nu \propto \nu^\alpha$, where $F_\nu$ is the integrated flux density at frequency $\nu$, and $\alpha$ is the spectral index. The spectral index uncertainties were derived using the following expression \citep{2016MNRAS.463.2997M}: 
\begin{equation}
{\alpha _{{\rm{err}}}} = \frac{1}{{{\rm{In}}\frac{{{v_1}}}{{{v_2}}}}}\sqrt {{{\left( {\frac{{{F_{1,err}}}}{{{F_1}}}} \right)}^2} + {{\left( {\frac{{{F_{2,err}}}}{{{F_2}}}} \right)}^2}} ,
	\label{eq:spix}
\end{equation}
where $v_{1,2}$ are the observed frequencies and $F_{1,2}$ represent the corresponding integrated flux densities. This accounts for the increased uncertainty in spectral indices derived from narrower frequency ranges. The flux density uncertainties in the TGSS ADR1 and NVSS surveys are approximately 10 percent \citep{2017A&A...598A..78I} and 5 percent \citep{1998AJ....115.1693C}, respectively. Using Equation \ref{eq:spix}, we estimate the spectral index uncertainty for the sources in our sample to be $\Delta\alpha$=0.05.

Columns (10) and (11) list the largest angular size (LAS) and the largest linear size (LLS), respectively. The LAS of each FR-I source was determined by measuring the maximum separation between boundary coordinates of the predicted mask obtained after applying a threshold of 3 times the local RMS level to the source's predicted bounding box in the FIRST image. As an example, the LAS of the source J0050+1036 is 66 arcsec, which is slightly greater than the length of its ridgeline shown in the upper panel of Figure~\ref{fig:ridgeline_example}. This discrepancy occurs because the ridgeline does not extend fully to the source's boundary at both ends. The 1400 MHz radio power ($P_{\rm 1400}$) of our FR-I sources, listed in Column (12), was calculated using the standard formula \citep[e.g.,][]{2009MNRAS.392..617D}: 
\begin{equation}
{P_{\rm 1400}} = \frac{{4{\rm \pi}{F_{\rm 1400}} {{({D_{L}})}^2}}}{{{{(1 + z)}^{(1 + \alpha^{\rm 1400}_{\rm 150})}}}} ,
	\label{eq:power}
\end{equation}
where $D_L$ is the luminosity distance to the FR-I source (in units of Mpc), $F_{\rm 1400}$ is the integrated flux density at 1400~MHz (in Jy), $z$ is the redshift of the source, $\alpha^{\rm 1400}_{\rm 150}$ (defined as $F_\nu \propto \nu^\alpha$) is the spectral index between 150~MHz and 1400~MHz. 


\subsection{Radio properties} \label{sec:radio_props}
\begin{figure}[ht!]
\centering
	\includegraphics[scale=0.5]{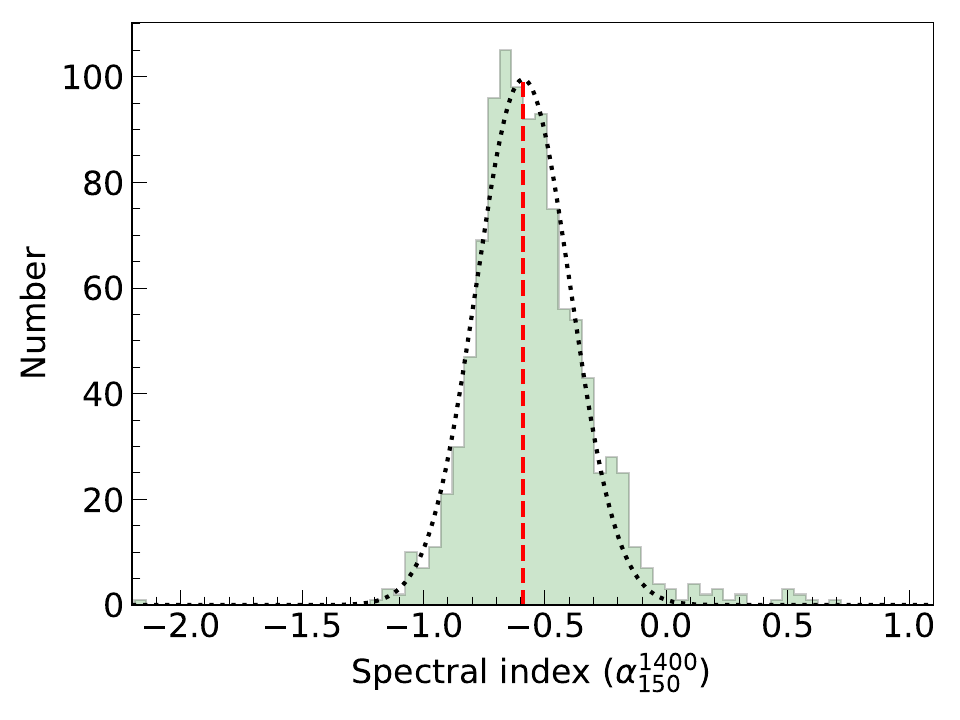}
    \caption{Distribution of spectral indexes ($\alpha_{\rm 150}^{\rm 1400}$) for the FR-I sample. The black dotted curve shows a Gaussian fit to the distribution (mean=-0.59$\pm$0.01, standard deviation=0.20), with the red dashed line marking the mean value.
    \label{fig:spix}}
\end{figure}

We measured spectral indexes for 947 FR-Is in our catalog. The remaining FR-Is were excluded from spectral index analysis because they were not detected (i.e., the measured region for each of these sources contains only NaN flux values) in the TGSS survey. Figure~\ref{fig:spix} shows the spectral index distribution, which ranges from -2.19 to +0.72. The median spectral index for the FR-I sample is $\alpha^{\rm 1400}_{\rm 150}$=-0.56, with a mean value of $\alpha^{\rm 1400}_{\rm 150}$=-0.58. This distribution is consistent with synchrotron emission from aged electron populations in FR-I lobes. The small offset (0.02) between the median and mean values indicates a near-symmetric distribution, suggesting negligible skewness from outliers or systematic errors. The flatter spectrum compared to the typical FR-I spectral index of -0.7 \citep[e.g.,][]{2010MNRAS.408.2261K} may result from our selection of core-dominated, bright objects in the FIRST survey, where the contribution of flat-spectrum cores reduces the integrated spectral steepness. Among them, 328 sources (34.6\%) are flat-spectrum objects ($\alpha>-$0.5), 618 (65.3\%) exhibit steep spectra ($-1.2\leq \alpha \leq-0.5$), and only one (0.1\%) qualifies as an ultra-steep spectrum source ($\alpha<-1.2$).

The LAS of the FR-Is range from 23 to 159 arcsec, with a mean and median LAS of 38 arcsec and 34 arcsec, respectively. The corresponding LLS of our FR-I sample spans from 2 to 561 kpc, exhibiting a mean LLS of 155 kpc and a median of 146 kpc. Notably, all FR-I sources in this study have LLS values below 700 kpc, which is the conventional threshold for classifying giant radio galaxies (GRGs) \citep[e.g.,][]{2018ApJS..238....9K}. 
This result is consistent with the established trend that FR-I sources generally exhibit smaller LLS values compared to FR-IIs, the latter of which can extend to significantly larger physical scales (reaching several Mpc in extreme cases). We note that the majority of GRGs exhibit FR-II morphology \citep[e.g.,][]{2024ApJS..273...30B}. The smaller size of FR-Is is likely attributable to enhanced jet disruption mechanisms, including (but not limited to): (1) stronger interaction with the interstellar medium, (2) increased entrainment of ambient material, and (3) the development of hydrodynamic instabilities in lower-power jets.

\begin{figure}[ht!]
    \centering
	\includegraphics[scale=0.5]{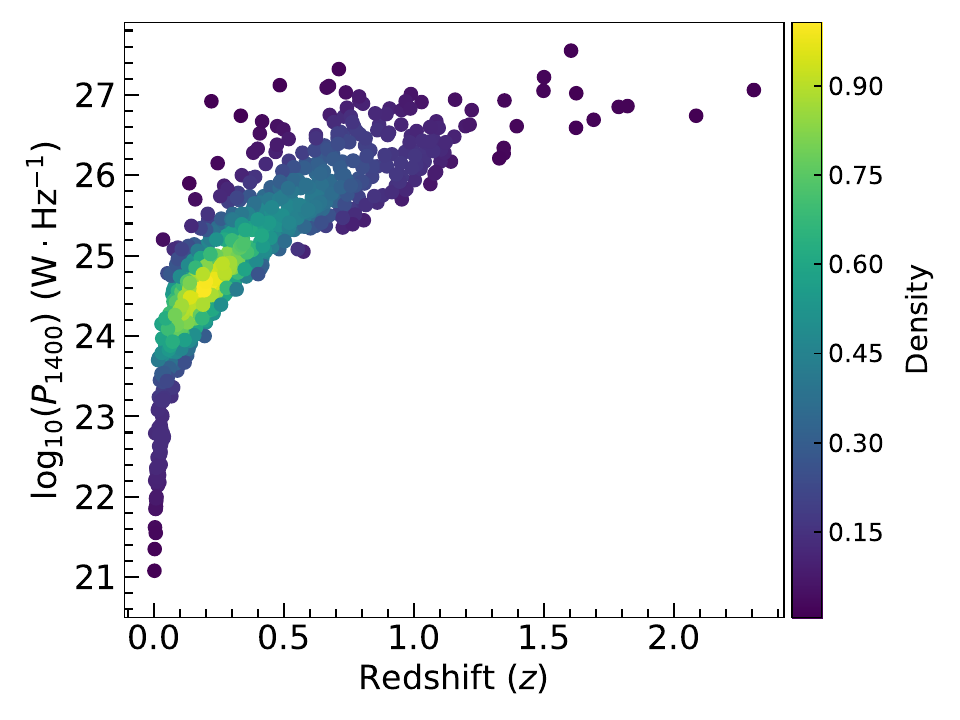}
    \caption{The logarithm of radio power at 1400 MHz (${\rm log}_{10}(P_{1400})$) as a function of redshift ($z$) for FR-I sources with available redshift and spectral index measurements. Color shading shows the 2D Gaussian kernel density estimate (KDE) of the source distribution in the $P_{1400}-z$ plane. The associated color bar encodes the relative source density (number of sources per unit area), with the `viridis' colormap ranging from low-density (purple) to high-density (yellow).  
    \label{fig:radio_P_L}}
\end{figure}

The radio power at 1400 MHz was successfully calculated for 793 FR-I sources for which
redshift and spectral index were available. Figure~\ref{fig:radio_P_L} presents the logarithm of radio power (${\rm log}_{10}(P_{1400})$) as a function of redshift. For the 793 FR-I sources analyzed, the 1400 MHz radio power spans ${\rm log}_{10}(P_{1400}/({\rm W}\,{\rm Hz}^{-1}))$ from 21.08 to 27.55, with mean and median values of 24.99 and 25.0, respectively.  
This indicates that our FR-I sample provides evidence that the standard FR break of ${\rm log}_{10}(P_{1400}/({\rm W}\,{\rm Hz}^{-1})) = 25$ \citep{1974MNRAS.167P..31F,1984ARA&A..22..319B} still exists. However, our sample also includes FR-I sources with radio power ${\rm log}_{10}(P_{1400}/({\rm W}\,{\rm Hz}^{-1})) > 25$. This presence of high-power FR-Is aligns with recent studies \citep[e.g.,][]{2017A&A...601A..81C,2019MNRAS.488.2701M} suggesting that the classical power threshold proposed by \citet{1974MNRAS.167P..31F} may not strictly divide the FR morphological classes.

\subsection{Hosts properties} 
\begin{figure}[ht!]
    \centering
	\includegraphics[scale=0.5]{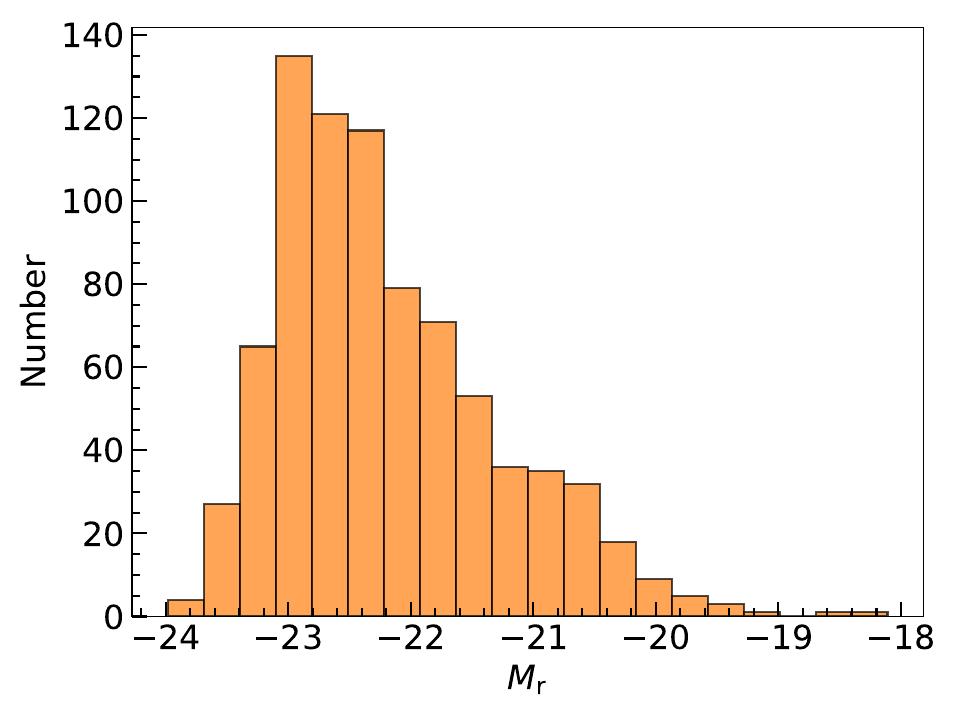}
    \caption{Distribution of the $r$-band absolute magnitudes ($M_r$) for 813 FR-I host galaxies.
    \label{fig:Mr}}
\end{figure}

We further collected and calculated the FR-I host galaxy properties, summarized for all 813 FR-I hosts in Table~\ref{tab:host_catalog}. Column (1) lists the short names from Table~\ref{tab:radio_catalog} for FR-Is with identified host galaxies. Column (2) provides the $r$-band apparent magnitude ($r$). Columns (3)-(6) list the $g-r$, $W1-W2$, $W2-W3$, and $z-W1$ color indices. All photometric measurements (magnitudes and colors) were taken from DESI LS DR10. Column (7) lists the $r$-band absolute magnitude ($M_r$), calculated using the following formula as defined in \citet{2024AJ....168...58D}:
\begin{equation}
M_r = r - 5 \log_{10}(D_L) - 25 - K_{\rm corr} - E_{\rm corr},
	\label{eq:Mr}
\end{equation}
where $K_{\rm corr}=-2.5{\rm log}_{10}(1+z_{\rm ref} )$ represents the $K$-correction, with $z_{\rm ref}$=0.1 as the reference redshift \citet{2024AJ....168...58D}. $E_{\rm corr}$ denotes the evolutionary correction, adopting the form from \citet{2014MNRAS.445.2125M} as $E_{\rm corr} = -Q_0(z - z_{\rm ref} )$, where we adopt $Q_0 = 0.97$. We assume pure luminosity evolution with no density evolution. All 813 FR-I host galaxies in our sample have been successfully analyzed to determine their $M_r$. The resulting $M_r$ distribution (see Figure~\ref{fig:Mr}) spans the range $-23.98 \leq M_r \leq -18.11$ mag, with mean and median values of $-22.18$ mag and $-22.40$ mag, respectively. The majority (98.4\%) of FR-I hosts have $M_r$ values within the approximate range of $-24 \lesssim M_r \lesssim -20$ mag, similar to the $M_r$ range reported for FR-Is by \citet{2017AA...598A..49C}.

\begin{deluxetable*}{lcccccccccccccccc}
\digitalasset
\tablewidth{0pt}
\setlength{\tabcolsep}{1pt}
\tablecaption{Hosts properties of the FR-I host galaxies. \label{tab:host_catalog}}
\tablehead{
   \colhead{Short name} & \colhead{r}  & \colhead{g-r} & \colhead{W1-W2} & \colhead{W2-W3} & \colhead{z-W1} & \colhead{$M_r$} & \colhead{${\rm log}_{10}(M_{\rm BH})$}  & \colhead{[OII]$\lambda5007$}  & \colhead{[OI]$\lambda$6300} & \colhead{[NII]$\lambda$6584} & \colhead{[SII]$\lambda$6716} & \colhead{${\rm H}_{\alpha}\lambda$6563} & \colhead{${\rm H}_{\beta}\lambda$4861}  & \colhead{[OIII] EW} & \colhead{EI} & \colhead{Class}\\
      \colhead{(J2000)} & \colhead{(mag)}  & \colhead{(mag)} & \colhead{(mag)} & \colhead{(mag)} & \colhead{(mag)} & \colhead{(mag)} & \colhead{($M_{\odot}$)}  & \colhead{($\times10^{-17}$)} & \colhead{($\times10^{-17}$)} & \colhead{($\times10^{-17}$)} & \colhead{($\times10^{-17}$)}  & \colhead{($\times10^{-17}$)}  & \colhead{($\times10^{-17}$)} & \colhead{($\mathring{\rm A}$)} & \colhead{}  & \colhead{}  \\
      & \colhead{}   &  \colhead{} & \colhead{} & \colhead{}  & \colhead{}  & \colhead{}  &  \colhead{}  & \colhead{(${\rm erg}\,{\rm s}^{-1}\,{\rm cm}^{-2}$)} & \colhead{(${\rm erg}\,{\rm s}^{-1}\,{\rm cm}^{-2}$)} & \colhead{(${\rm erg}\,{\rm s}^{-1}\,{\rm cm}^{-2}$)} & \colhead{(${\rm erg}\,{\rm s}^{-1}\,{\rm cm}^{-2}$)} & \colhead{(${\rm erg}\,{\rm s}^{-1}\,{\rm cm}^{-2}$)}  & \colhead{(${\rm erg}\,{\rm s}^{-1}\,{\rm cm}^{-2}$)}  & \colhead{} & \colhead{}  & \colhead{}  
}
\colnumbers
\startdata
J0001+1250 & 16.43 &1.36 &-0.38 &--& -0.03 &-23.33 &8.79 &--& 12.92 &--&--&1.97 &--&--&--&--\\
J0003+0709 &17.57 &1.32 &-0.36 &--&-0.13&-22.27&--&--&--&--&--&--&--&--&--&-- \\
J0003+0134 &22.48 &0.29 &0.71 &3.99 &1.53 &-20.58&--&--&--&--&--&--&--&--&--&HERG \\
J0012-0723 &17.75 &1.62 &-0.48 &-0.51 &0.33 &-23.04 &8.76 &8.05 &8.66 &--&0.45 &6.96 &--&--&--&LERG \\
J0012-0857 &18.24 &1.7 &-0.32 &0.42 &0.47 &-22.66 &8.51 &52.94 &10.07 &60.39 &22.08 &38.12 &1.23 &--&1.84 &HERG \\
J0015-0653 &20.21 &1.93 &-0.58 &0.01 &1.18 &-22.01 &8.84 &3.22&--&--&--&--&6.6 &--&--&LERG \\
J0017-0512 &17.57 &0.18 &0.36 &0.87 &1.83 &-22.47 &--&553.48 &34.06 &38.47 &52.54 &392.44 &130.14 &--&1.61 &HERG \\
J0018+1117 &18.04 &1.13 &-0.46 &-2.71 &-0.02 &-20.86 &8.16 &5.7 &4.82 &16.2 &8.83 &7.83 &0.93 &0.46 &0.73 &LERG \\
J0019+1155 &19.65 &1.62 &-0.34 &1.37 &0.84 &-21.89 &8.42 &11.35 &6.36 &14.91 &0.85 &17.93 &3.25 &--&1.16 &HERG \\
J0026-0556 &21.6 &0.43 &0.29 &-0.12 &1.98 &-21.99&--&--&--&--&--&--&--&--&--&LERG\\
J0027-0204 &17.21 &1.17 &-0.23 &-1.21 &0.21 &-22.81&--&--&--&--&--&--&--&--&--&LERG\\
J0031+1212&17.23 &1.29 &-0.54 &2.71 &-0.01 &-22.25 &8.69&--&--&10.82 &5.7 &7.78 &--&--&--&--\\
J0034-0638 &21.74 &0.77 &-0.27 &1.8 &1.54 &-21.29&--&--&--&--&--&--&--&--&--&-- \\
J0038-0207&17.75 &0.95 &0.34 &0.77 &1.28 &-22.23 &--&626.45 &224.28 &168.02 &123.34 &1727.46 &279.21 &--&1.37 &HERG\\
J0039-1032 &16.25 &1.02 &-0.48&--&0.02&-22.53 &8.44 &9.33 &--&7.18 &14.94 &11.13&--&--&--&--\\
J0042+1509 &21.66 &0.84 &0.29 &1.2 &2.75 &-21.7&--&--&--&--&--&--&--&--&--&LERG\\
J0043-1039 &15.56 &1.04 &-0.47 &-1.76 &0.02 &-23.2 &8.78 &15.81 &6.89 &38.57 &24.78 &25.09 &--&--&--& LERG\\
J0043-0459&16.68 &1.36 &-0.35 &-2.23 &0.12 &-23.11 &8.67&--&--&66.09&--&11.67&--&--&--&LERG\\
J0043+1355&16.76&1.14&-0.46&-3.63&0.02&-22.05&8.39&9.21&6.68&22.37&--&7.15&--&--&--&LERG\\
J0050+1036	&13.33	&0.94	&-0.58	&-1.03	&-0.31 &	-22.8 &--&--&--&--&--&--&--&--&--&	LERG \\
\enddata
\tablecomments{Column (1): short name (JHHMM+DDMM). Column (2): $r$-band apparent magnitudes (in mag) from DESI LS DR10. Columns (3) to (6): color indices from DESI LS DR10 comprising $g-r$, $W1-W2$, $W2-W3$, and $z-W1$ (in mag). Column (7): $r$-band absolute magnitudes in mag. Column (8): logarithmic black hole masses in $M_\odot$. Columns (9) to (14): emission-line integrated fluxes (in units of $10^{-17}\,{\rm erg\,s^{-1}\,cm^{-2}}$) for [OII]$\lambda5007$, [OI]$\lambda6300$, [NII]$\lambda6584$, [SII]$\lambda6716$, H$\alpha$$\lambda6563$, and H$\beta$$\lambda4861$. Column (15): the equivalent width (EW) of [OIII]. Column (16): the emission index (EI). Column (17): high-/low-excitation classification results. The full version of Table \ref{tab:host_catalog} is available at: \url{https://drive.google.com/file/d/1nwSFau1VLeJtVpQTrcYbZ8A_QQvS8WFf/view?usp=sharing}.}
\end{deluxetable*}

\begin{figure}[ht!]
    \centering
	\includegraphics[scale=0.5]{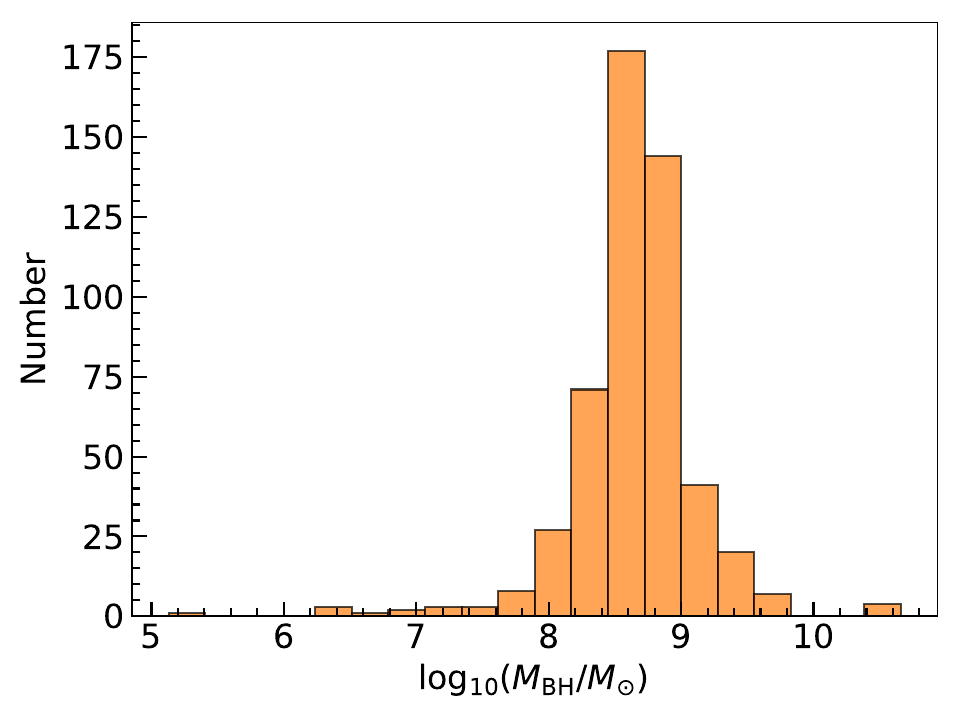}
    \caption{Distribution of the logarithm black hole masses ($M_{\rm BH}$) for 512 FR-I host galaxies.
    \label{fig:Mbh}}
\end{figure}

The black hole masses ($M_{\rm BH}$) of our FR-I host galaxies, listed in column (8) of Table \ref{tab:host_catalog}, were estimated using the $M-\sigma_{*}$ relation from \citet{2002ApJ...574..740T} as follows:

\begin{equation}
{\log _{10}}({M_{\rm BH}}/{M_ \odot }) = 8.13 + 4.02{\log _{10}}({\sigma _*}/{\sigma _0}) ,
	\label{eq:m-sigma}
\end{equation}
where $\sigma _*$ represents the stellar velocity dispersion, and $\sigma _0$=200 km/s. The $\sigma _*$ 
for our FR-I hosts were obtained directly from the value-added catalog of DESI DR1, which were derived using the FastSpecFit software \citep{2023ascl.soft08005M}, and the SDSS DR17 spectroscopic object catalog. We successfully computed $M_{\rm BH}$ for 512 FR-I hosts. The distribution of the logarithm black hole masses ($\log_{10}(M_{\rm BH}/M_{\odot})$) is presented in Figure~\ref{fig:Mbh}, with values spanning $5.13 \leq \log_{10}(M_{\rm BH}/M_{\odot}) \leq 10.66$. The sample shows a mean (median) $\log_{10}(M_{\rm BH}/M_{\odot})$ of 8.64 (8.67), with 97.9\% of hosts concentrated in the range $7.0 \lesssim \log_{10}(M_{\rm BH}/M_{\odot}) \lesssim 9.9$.

Classifying radio galaxies into HERGs or LERGs by their optical emission line ratios facilitates studying their formation and evolution. We extracted the integrated fluxes of the primary optical emission lines, including [OIII]$\lambda$5007, [OI]$\lambda$6300, [NII]$\lambda$6584, [SII]$\lambda$6716, $\mathrm{H}_\alpha\lambda6563$, and $\mathrm{H}_\beta\lambda4861$, directly from the value-added catalog of DESI DR1 and the spectroscopic object catalog of SDSS DR17, which are listed in columns (9)-(14) in Table~\ref{tab:host_catalog}. The emission-line fluxes in DESI DR1 were measured using FastSpecFit \citep{2023ascl.soft08005M}, which performs simultaneous modeling of the stellar continuum and emission lines while incorporating broadband photometry. The emission-line fluxes in the SDSS DR17 spectroscopic object catalog were derived using Gaussian fitting after stellar-continuum subtraction. We also obtained the equivalent width (EW) of the [OIII]$\lambda$5007 emission line from the same datasets, listed in column (15). Based on the emission line information, we computed the excitation index (EI) using the formula from \citet{2010A&A...509A...6B}, with the results listed in column (16):  
\begin{equation}
\begin{split}
{\rm EI} = \log_{10}\left( \frac{\rm [OIII]}{{\rm H}_\beta} \right) - \frac{1}{3}
\Bigg[\log_{10}\left( \frac{\rm [NII]}{{\rm H}_\alpha} \right) + \\ 
\log_{10}\left( \frac{\rm [SII]}{{\rm H}_\alpha} \right) + \log_{10}\left( \frac{\rm [OI]}{{\rm H}_\alpha} \right) \Bigg].
\end{split}
\end{equation}

As the first step of our analysis, we classify FR-I hosts into HERGs and LERGs using the method of \citet{2010A&A...509A...6B}: if EI$>$0.95, the host galaxies are classified as HERGs; otherwise, they are classified as LERGs. Then, if the EI is not available, we use the flux ratio classification method from \citet{2012MNRAS.421.1569B}, which reports [OIII] EW$> 5$ Å and optical flux ratio [OIII]$\lambda5007/{\rm H}\alpha\lambda6536 \geq 1$ for HERGs. Furthermore, radio galaxies can also be classified into HERGs and LERGs using WISE mid-IR colors \citep[e.g.,][]{2017MNRAS.469.2886D,2020A&A...642A.153D}: HERGs are defined by W1-W2$>$0.5 and W2-W3$<$5.1, while LERGs satisfy W1-W2$<$0.5 and 0$<$W2-W3$<$1.6. After this series of classifications, we successfully categorized 630 FR-I hosts, among which 571 (90.6\%) are LERGs and 59 (9.4\%) are HERGs. The classification results are listed in column (17). This finding aligns with previous reports that most FR-Is are LERGs \citep[e.g.,][]{2009ApJ...699L..43S}.

\begin{figure}[ht!]
    \centering
	\includegraphics[scale=0.5]{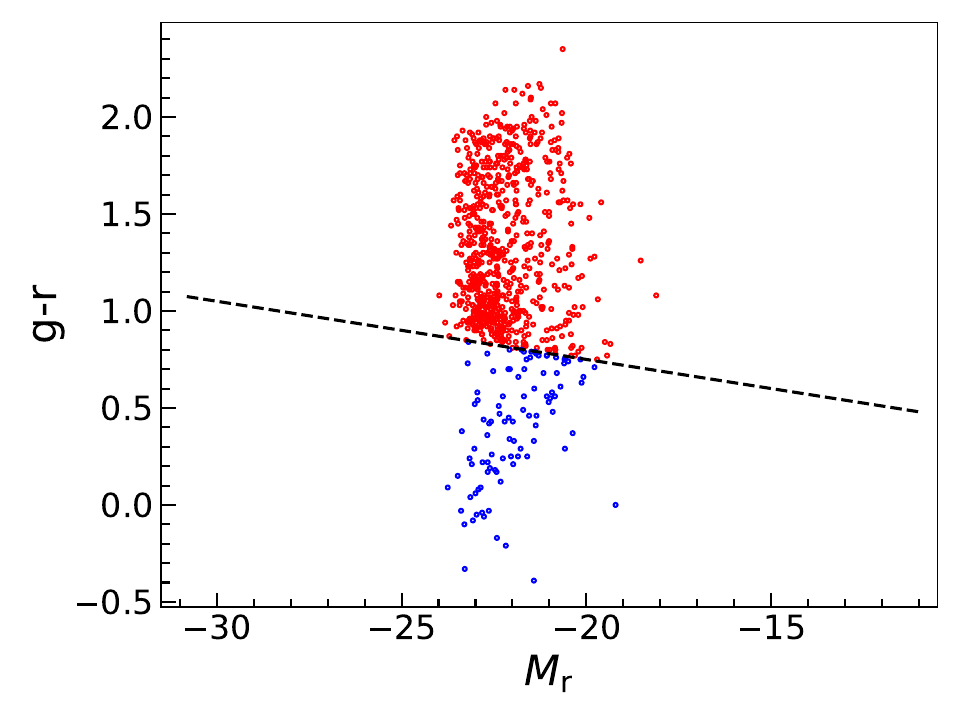}
    \caption{Distribution of the $g-r$ color as a function of $M_r$ for the FR-I host galaxies. The black dashed line represents the separation method from \citet{2006MNRAS.366....2W}, which divides the hosts into red and blue galaxies.
    \label{fig:g-r_Mr}}
\end{figure}

Figure~\ref{fig:g-r_Mr} shows the distribution of the $g-r$ color as a function of $M_r$ for the FR-I host galaxies. Photometric $g-r$ values were successfully extracted from the DESI LS DR10 for all 813 FR-I hosts in our sample. Following the classification scheme of \citet{2006MNRAS.366....2W}, we separate the host galaxies into red and blue populations using the black dashed line in Figure~\ref{fig:g-r_Mr}. Hosts lying above this demarcation line are classified as red galaxies (red solid circle), while those below are considered blue galaxies (blue solid circle). Our analysis reveals that 88.1\% of FR-I hosts (716) exhibit elevated $g-r$ colors, placing them within the red sequence; the remaining 97 hosts (11.9\%) are identified as blue galaxies. We found that the majority (84.9\%) of blue host galaxies exhibit high radio power (${\rm log}_{10}(P_{1400}/({\rm W}\,{\rm Hz}^{-1}))>25$), suggesting that they may represent a population of `quenching' galaxies.

\begin{figure*}[ht!]
    \centering
	\includegraphics[scale=0.8]{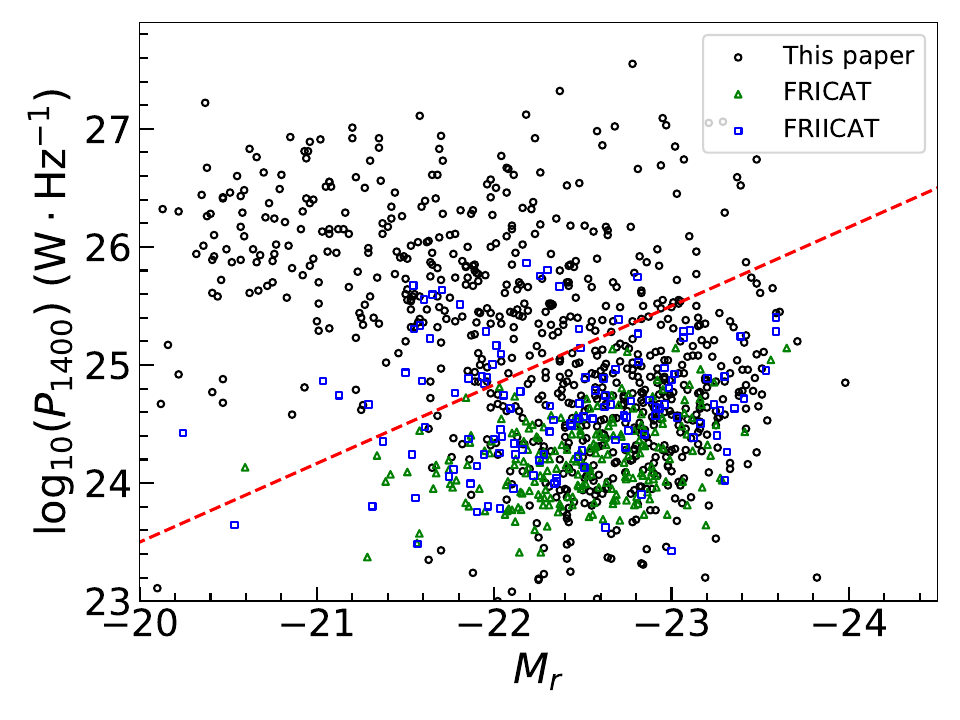}
    \caption{The 1400 MHz radio power vs. $r$-band absolute magnitude ($M_r$). The power division line (the red dashed line) is from \citet{1996AJ....112....9L}. The FR-Is from the \citep[FRICAT;][]{2017AA...598A..49C} are marked with green open triangles. The FR-IIs from the \citep[FRIICAT;][]{2017A&A...601A..81C} are marked with blue open squares. The black open circles represent the FR-Is from this paper.   
   \label{fig:P_Mr}}
\end{figure*}

\subsection{The separation between FR classes}
Figure~\ref{fig:P_Mr} shows the correlation between 1400 MHz radio power and $M_r$, combining our sample (black open circles) with FR-Is (green open triangles) \citep[FRICAT;][]{2017AA...598A..49C}) and FR-IIs (blue open squares) \citep[FRIICAT;][]{2017A&A...601A..81C}) for comparative analysis. The data points representing FR-Is from this paper (black open circles) are spread across a wide range of $M_r$ and $\log_{10}(P_{1400})$ values. In the region where $M_r$ is approximately between -21.5 and -23.8 mag, these data points show a relatively higher concentration. This could imply that within this $M_r$ range, the FR-Is studied in this paper exhibit certain characteristic radio power properties. Compared to data from other sources (FRICAT and FRIICAT), our data points have unique distributions in some areas. Data points from this paper (FR-Is) are found on both sides of the power division line \citep{1996AJ....112....9L}. When compared to FRICAT data points (green open triangles), differences in their proximity to the power division line can be noticed. This suggests that the selection of FR-I objects differed between the two studies and further demonstrates that the power division may not strictly separate the FR morphological classes. FR-IIs (blue open squares) from the FRIICAT are also distributed on both sides of the line. This indicates that both our FR-Is and the FR-IIs in FRIICAT, similar to those reported by \citet{2019MNRAS.488.2701M}, fail to exhibit a strict separation based on the power division line. For example, in the region where $M_r$ is around -22.5 mag, some FR-IIs are found on the lower radio power side (below the line), which is comparable to the distribution of some of our FR-Is in that area. This shared characteristic of crossing the power division line further emphasizes the complexity of the relationship between radio power, $M_r$, and FR morphological classes, and suggests that factors other than a simple power division may be at play in determining the properties of these sources.

\section{Conclusions} \label{Sec:Conc}

In this paper, we built a catalog of 971 FR-I radio galaxies identified from the VLA FIRST survey, 845 of which were newly discovered. Our catalog is the largest and most comprehensive to date, representing an approximately fourfold increase in sample size over previous compilations from the VLA FIRST survey. It will enable statistical studies of FR-I populations with unprecedented sample size and reliability, while serving as a critical training dataset for future artificial intelligence-based morphological classification of radio galaxies. Our identification pipeline initially detected 12,501 FR-I candidates from 946,366 FIRST images using a deep learning-based radio source detector (RGCMT model), with final classifications determined through the proposed ridgeline flux density distribution-based identification method. This method, which is based on the continuous variation of flux density in FR-I jets, is more reliable than the discrete-point determination approach proposed by \citet{2019MNRAS.488.2701M} (wherein distances are calculated to emission peaks relative to half the total extents). Moreover, compared to traditional visual inspection using multi-band overlaid images, our numerical statistical analysis significantly reduces subjective misjudgments arising from human factors. The ridgeline flux density distribution method we propose will provide a novel toolset for studying the evolution of radio galaxies. For example, by enabling statistical analysis of ridgeline position angles across different epochs, this method makes it possible to investigate the kinematic evolution of radio jets \citep{2025ApJ...987..169S}.

To complement the morphological analysis, host galaxies for 813 FR-Is were identified through cross-matching with the DESI LS DR10 catalog, followed by visual inspection to determine the host candidates and exclude star candidates.
The FR-Is in this catalog exhibit radio luminosities of $1.20 \times 10^{21} \leq P_{\rm 1400} \leq 3.55 \times 10^{27}\,{\rm W\,Hz}^{-1}$ and redshifts of up to 2.307. Their hosts have $M_r$ values predominantly between -24 and -20 mag, with $M_{\rm BH}$ spanning $7.0 \lesssim \log_{10}(M_{\rm BH}/M_{\odot}) \lesssim 9.9$. Moreover, 88.1 per cent of the hosts are red galaxies. 
The steep spectral index between 150 and 1400 MHz
was found for 65.3\% of the FR-Is. All identified FR-Is exhibit LLS below 700 kpc, confirming their exclusion from the GRG population (LLS$\geq$700 kpc). This size distribution is consistent with previous studies showing that FR-Is rarely evolve into GRGs. We classified FR-I hosts into HERG and LERG using optical emission-line ratios and mid-infrared colors, among which 571 hosts were LERGs and 59 hosts were HERGs. The results of the correlation between 1400 MHz radio power and $r$-band absolute magnitude for our FR-Is do not support the strict separation of FR classes by power division lines. The aforementioned physical properties and analyses of the FR-Is will serve as valuable references for extragalactic radio source studies.


\begin{acknowledgments}
This work was supported by the National SKA Program of China (2022SKA0120101, 2022SKA0130100, 2022SKA0130104), the National Natural Science Foundation of China (Nos. 12103013), the Foundation of Science and Technology of Guizhou Province (Nos. (2021)023), the Foundation of Guizhou Provincial Education Department (Nos. KY(2021)303, KY(2020)003, KY(2023)059), and the Guangxi Universities Engineering Research Center for Optoelectronic Information Technology. X.Y. was supported by the National Science Foundation of China (12103076, 12233005), the National Key R\&D Program of China (2020YFE0202100), the Shanghai Sailing Program (21YF1455300), and the China Postdoctoral Science Foundation (2021M693267). 
T.K.S. acknowledges the funding support from the National SKA Program of China under Nos. 2022SKA0110100 and 2022SKA0110101.

This research used data obtained with the Dark Energy Spectroscopic Instrument (DESI). DESI construction and operations is managed by the Lawrence Berkeley National Laboratory. This material is based upon work supported by the U.S. Department of Energy, Office of Science, Office of High-Energy Physics, under Contract No. DE-AC02-05CH11231, and by the National Energy Research Scientific Computing Center, a DOE Office of Science User Facility under the same contract. Additional support for DESI was provided by the U.S. National Science Foundation (NSF), Division of Astronomical Sciences under Contract No. AST-0950945 to the NSF’s National Optical-Infrared Astronomy Research Laboratory; the Science and Technology Facilities Council of the United Kingdom; the Gordon and Betty Moore Foundation; the Heising-Simons Foundation; the French Alternative Energies and Atomic Energy Commission (CEA); the National Council of Humanities, Science and Technology of Mexico (CONAHCYT); the Ministry of Science and Innovation of Spain (MICINN), and by the DESI Member Institutions: www.desi.lbl.gov/collaborating-institutions. The DESI collaboration is honored to be permitted to conduct scientific research on I’oligam Du’ag (Kitt Peak), a mountain with particular significance to the Tohono O’odham Nation. Any opinions, findings, and conclusions or recommendations expressed in this material are those of the author(s) and do not necessarily reflect the views of the U.S. National Science Foundation, the U.S. Department of Energy, or any of the listed funding agencies.

The DESI Legacy Imaging Surveys consist of three individual and complementary projects: the Dark Energy Camera Legacy Survey (DECaLS), the Beijing-Arizona Sky Survey (BASS), and the Mayall z-band Legacy Survey (MzLS). DECaLS, BASS and MzLS together include data obtained, respectively, at the Blanco telescope, Cerro Tololo Inter-American Observatory, NSF’s NOIRLab; the Bok telescope, Steward Observatory, University of Arizona; and the Mayall telescope, Kitt Peak National Observatory, NOIRLab. NOIRLab is operated by the Association of Universities for Research in Astronomy (AURA) under a cooperative agreement with the National Science Foundation. Pipeline processing and analyses of the data were supported by NOIRLab and the Lawrence Berkeley National Laboratory (LBNL). Legacy Surveys also uses data products from the Near-Earth Object Wide-field Infrared Survey Explorer (NEOWISE), a project of the Jet Propulsion Laboratory/California Institute of Technology, funded by the National Aeronautics and Space Administration. Legacy Surveys was supported by: the Director, Office of Science, Office of High Energy Physics of the U.S. Department of Energy; the National Energy Research Scientific Computing Center, a DOE Office of Science User Facility; the U.S. National Science Foundation, Division of Astronomical Sciences; the National Astronomical Observatories of China, the Chinese Academy of Sciences and the Chinese National Natural Science Foundation. LBNL is managed by the Regents of the University of California under contract to the U.S. Department of Energy. The complete acknowledgments can be found at https://www.legacysurvey.org/acknowledgment/.

Funding for the Sloan Digital Sky 
Survey IV has been provided by the 
Alfred P. Sloan Foundation, the U.S. 
Department of Energy Office of 
Science, and the Participating 
Institutions. 

SDSS-IV acknowledges support and 
resources from the Center for High 
Performance Computing  at the 
University of Utah. The SDSS 
website is www.sdss4.org.

SDSS-IV is managed by the 
Astrophysical Research Consortium 
for the Participating Institutions 
of the SDSS Collaboration including 
the Brazilian Participation Group, 
the Carnegie Institution for Science, 
Carnegie Mellon University, Center for 
Astrophysics | Harvard \& 
Smithsonian, the Chilean Participation 
Group, the French Participation Group, 
Instituto de Astrof\'isica de 
Canarias, The Johns Hopkins 
University, Kavli Institute for the 
Physics and Mathematics of the 
Universe (IPMU) / University of 
Tokyo, the Korean Participation Group, 
Lawrence Berkeley National Laboratory, 
Leibniz Institut f\"ur Astrophysik 
Potsdam (AIP),  Max-Planck-Institut 
f\"ur Astronomie (MPIA Heidelberg), 
Max-Planck-Institut f\"ur 
Astrophysik (MPA Garching), 
Max-Planck-Institut f\"ur 
Extraterrestrische Physik (MPE), 
National Astronomical Observatories of 
China, New Mexico State University, 
New York University, University of 
Notre Dame, Observat\'ario 
Nacional / MCTI, The Ohio State 
University, Pennsylvania State 
University, Shanghai 
Astronomical Observatory, United 
Kingdom Participation Group, 
Universidad Nacional Aut\'onoma 
de M\'exico, University of Arizona, 
University of Colorado Boulder, 
University of Oxford, University of 
Portsmouth, University of Utah, 
University of Virginia, University 
of Washington, University of 
Wisconsin, Vanderbilt University, 
and Yale University.

This research uses services or data provided by the Astro Data Lab, which is part of the Community Science and Data Center (CSDC) Program of NSF NOIRLab. NOIRLab is operated by the Association of Universities for Research in Astronomy (AURA), Inc. under a cooperative agreement with the U.S. National Science Foundation.

This research has made use of the NASA/IPAC Extragalactic Database (NED), which is funded by the National Aeronautics and Space Administration and operated by the California Institute of Technology.

\end{acknowledgments}

\begin{contribution}


BL conceived the initial research concept, wrote the manuscript, reviewed samples, and implemented the analysis methods.
XY is responsible for sample review, manuscript editing, data analysis, and explaining the results.
WX performed sample review and implemented the analysis method. He also edited the manuscript.
TKS edited the manuscript and data analysis.
YZ is responsible for sample review and manuscript editing. 
DL developed the software and maintains the project's GitHub repositories.
ZL is responsible for data analysis.
YL edited the manuscript.
RZ is responsible for sample review.

\end{contribution}

%
\facilities{CDS, VLA, WISE, Sloan, CTIO:2MASS, Astro Data Lab}

\software{The work carried out in this paper made use of the
following software packages and tools: Astropy 
 \citep{2013A&A...558A..33A,2018AJ....156..123A}, APLpy \citep{2012ascl.soft08017R}, Matplotlib \citep{2007CSE.....9...90H}, Astroquery \citep{2019AJ....157...98G}, Numpy \citep{2020Natur.585..357H}, Pandas \citep{mckinney-proc-scipy-2010,reback2020pandas}, scikit-image \citep{van2014scikit}, SciPy \citep{virtanen2020scipy}, SAOImage
DS9 \citep{2003ASPC..295..489J}, TOPCAT \citep{2005ASPC..347...29T}}. 

\bibliography{sample701}{}
\bibliographystyle{aasjournalv7}



\end{document}